\documentclass[11pt,a4paper]{article}
\pdfoutput=1

\usepackage{amsfonts}
\usepackage{mathrsfs}
\usepackage[dvipdfmx]{color}
\RequirePackage{ifpdf}
\ifpdf
  \usepackage[pdftex]{graphicx}
\else
  \usepackage[dvipdfmx]{graphicx}
\fi
\usepackage{jheppub_kim}
\usepackage{amssymb}
\usepackage{color}
\usepackage{amsmath}

\usepackage{amsfonts}
\usepackage{epstopdf}
\usepackage{slashed}
\usepackage{extarrows}

\usepackage{multirow, graphicx,amssymb,url,mathrsfs,amsmath}
\usepackage{wrapfig,boxedminipage,subfigure,epsfig}
\usepackage{amsxtra,amstext,latexsym,dsfont,amsfonts}
\usepackage{color,eucal}
\usepackage[dvipsnames]{xcolor}
\usepackage{float}

\renewcommand{\a}{\alpha}      \renewcommand{\b}{\beta}
      \renewcommand{\k}{\kappa}

\title{Entanglement after Quantum Quenches in Lifshitz Scalar Theories}

\author[a]{Keun-Young~Kim,}
\author[a]{Mitsuhiro~Nishida,}
\author[b,c]{Masahiro~Nozaki,}
\author[a]{Minsik~Seo,}
\author[d]{Yuji~Sugimoto,}
\author[e]{Akio~Tomiya}

\emailAdd{fortoe@gist.ac.kr}
\emailAdd{mnishida@gist.ac.kr}
\emailAdd{masahiro.nozaki@riken.jp}
\emailAdd{tjalstlr23@gist.ac.kr}
\emailAdd{sugimoto@ustc.edu.cn}
\emailAdd{akio.tomiya@riken.jp}

\affiliation[a]{School of Physics and Chemistry, Gwangju Institute of Science and Technology, Gwangju 61005, Korea}
\affiliation[b]{iTHEMS Program, RIKEN, Wako, Saitama 351-0198, Japan}
\affiliation[c]{Berkeley Center for Theoretical Physics, Department of Physics, University of California, Berkeley, CA 94720, USA}		
\affiliation[d]{Interdisciplinary Center for Theoretical Study, University of Science and Technology of China, Hefei, Anhui 230026, China}
\affiliation[e]{RIKEN/BNL Research center, Brookhaven National Laboratory,  Upton, NY, 11973, USA}

\abstract{
We study the time evolution of the entanglement entropy after quantum quenches in {Lifshitz free scalar theories}, with the dynamical exponent $z>1$, by using the correlator method. For quantum quenches we consider two types of time-dependent mass functions:  end-critical-protocol (ECP) and cis-critical-protocol (CCP). In both cases, {at early times} the entanglement entropy is independent of the subsystem size.  After a critical time ($t_c$), the entanglement entropy starts depending on the {subsystem} size significantly. This critical time $t_c$ for $z = 1$ in the fast ECP and CCP has been explained well by the fast quasi-particle of the quasi-particle picture. However, we find that for $z > 1$ this explanation does not work and $t_c$ is delayed.
We explain why $t_c$ is delayed for $z>1$ based on the quasiparticle picture: in essence, it is due to the competition between the fast and slow quasiparticles. {At late times}, in the ECP, the entanglement entropy slowly increases while, in the CCP, it is oscillating with a well defined period 
by the final mass scale, independently of $z$. We give an interpretation of this phenomena by the correlator method. As $z$ increases, the entanglement entropy increases, which can be understood by long-range interactions due to $z$. }

\begin{document}

\begin{flushright} USTC-ICTS-19-14\\ RIKEN-iTHEMS-Report-19\end{flushright} 

\maketitle

\section{Introduction}
\begin{figure}[]
 \centering
     \subfigure[End-critical-protocol (ECP)]
     {{\includegraphics[width=5 cm]{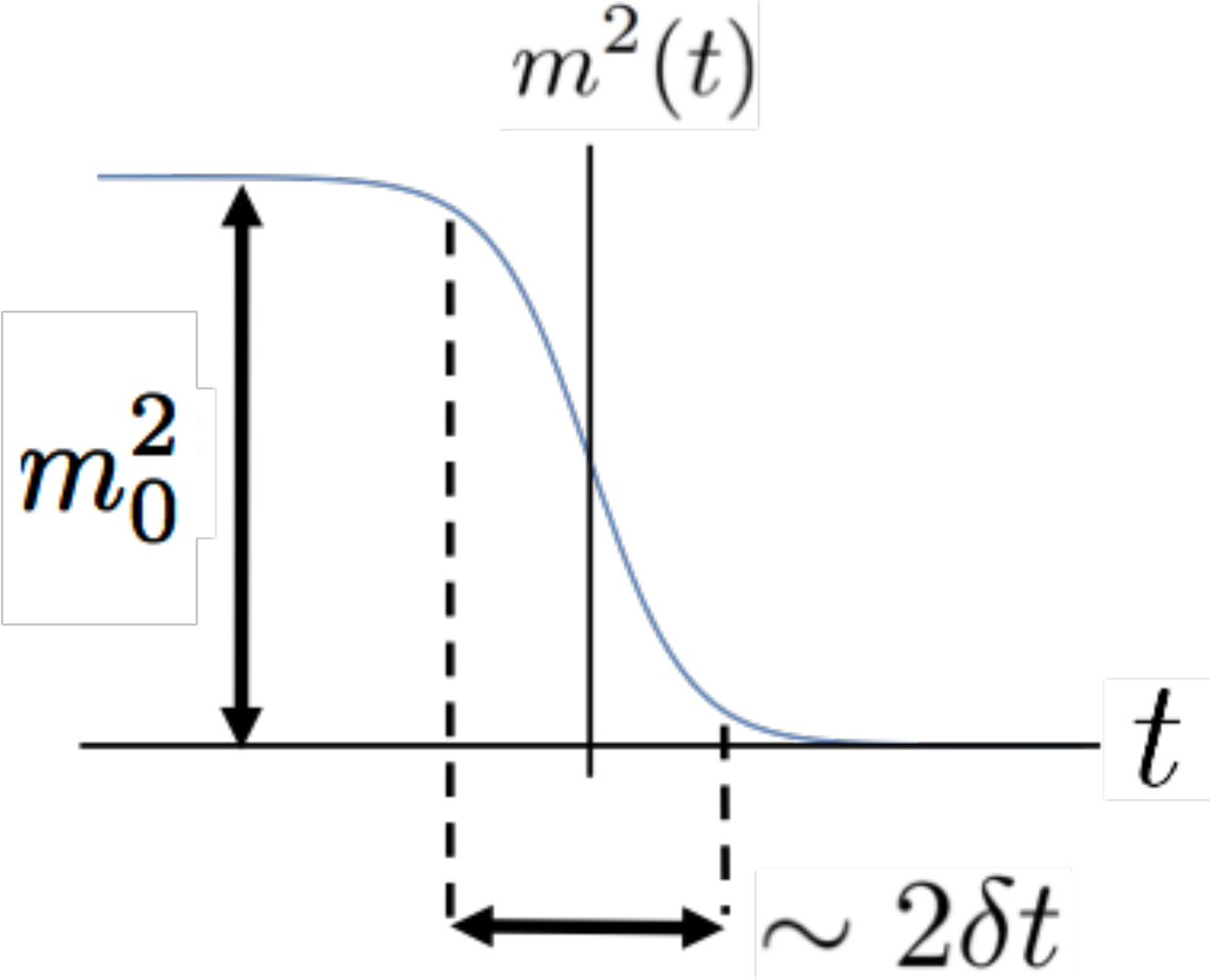} \label{fig1a}} }  \ \ \ \ \ \ \ \
     \subfigure[Cis-critical-protocol (CCP)]
     {{\includegraphics[width=5 cm]{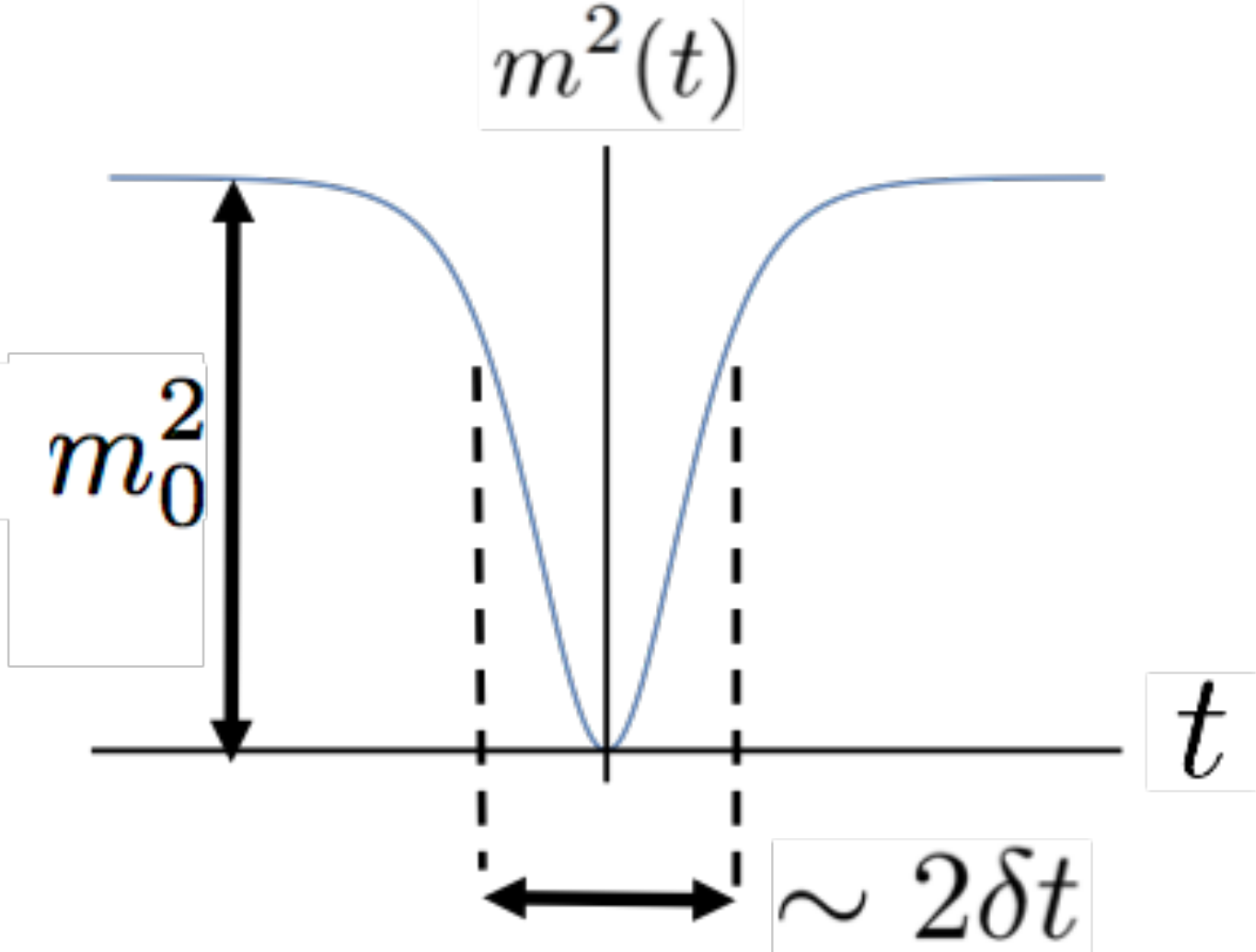} \label{fig1b}}}
           \caption{Schematic descriptions of the mass potential $m^2(t)$ in the ECP and the CCP. We will explain the ECP and CCP in more details in section \ref{sup}. }\label{quench}
\end{figure}

Time evolution of non-equilibrium systems is an important subject in physics such as thermalization processes of quantum many body systems and black hole formation (see reviews \cite{Gogolin:2016hwy, Page:2004xp}). 
One well-studied protocol to describe the {time} evolution process of non-equilibrium systems is the quantum quench with a time dependent Hamiltonian (see for example \cite{Calabrese:2006rx, Calabrese_2016,  Calabrese:2007rg} and Figure \ref{quench} in this paper).
In this case, one can calculate time evolution of the system
and obtain insights {on} the time evolution through
a measure of entanglement.

Typical choices of the time-dependent mass potentials are the ones in {the} end-critical-protocol (ECP) and cis-critical-protocol (CCP) \cite{Chandran_2012}. In the ECP, the mass potential is nonzero at early times and approaches to zero at late times as shown in Figure \ref{fig1a}. On the other hand, in the CCP, the mass potential is nonzero at $t\to\pm\infty$ and becomes zero at $t=0$  as shown in Figure \ref{fig1b}. {The scaling property,} the time evolution of correlation functions, entanglement measures, and complexity in the ECP and CCP were studied in \cite{ Das:2014jna, Das:2014hqa, Das:2015jka, Das:2016lla,  Chandran_2012, Caputa:2017ixa, Alves:2017fjk, Nishida:2017hqd,  Alves:2018qfv,  Camargo:2018eof, Fujita:2018lfj, MohammadiMozaffar:2019gpn}.

When we consider the time evolution of a pure state {due to a unitary time evolution operator}, the {density matrix} cannot become a {mixed} state. However,  the {reduced density matrix} of the total system can be the mixed states.
After a sufficient time, this subsystem may show the properties of {thermodynamic equilibrium}. {A measure to study these properties is the entanglement entropy, which is defined by von Neumann entropy for a reduced density matrix:}
\begin{align}
S_A=-\textrm{Tr}_A\rho_A\log \rho_A\,,
\end{align} 
where $\rho_A$ is the reduced density matrix of the subsystem $A$. If the entanglement entropy $S_A$ behaves as a {thermodynamic entropy} of an {equilibrium} state, {one can interpret the subsystem $A$ as  {thermodynamic equilibrium}}.

The time evolution of the entanglement entropy for an interval in {2 dimensional  conformal field theories (CFT)} in a sudden quench, which is a protocol that the mass in the Hamiltonian is suddenly changed at $t=0$, is well described by the quasiparticle picture (see, for example, \cite{Calabrese:2007rg, Peschel_2009, Calabrese:2005in,   Cotler:2016acd} and Figure \ref{quasip} in this paper). 
The basic idea is {as follows:} i) by a sudden quench, the quasiparticle pairs are generated ii) these quasiparticles {contribute to the change of  entanglement entropy after the quench.}
For example, if the final mass after the quench is small enough, the maximum propagation speed of the quasiparticles is approximately the speed of light, so the entanglement entropy starts depending on subsystem size $l$ from $t\sim\frac{l}{2}$ when $l$ is large compared with the initial correlation length.\footnote{This will be explained in more detail in section \ref{IECP}.} This result agrees with the analysis of {2 dimensional  CFT} in a sudden quench. The quasiparticle formula in the sudden quench including the quasiparticles with various group velocities was  studied in \cite{ Alba:2017aa, Alba:2017lvc}.

Instead of systems with the Lorentz symmetry, one can consider the Lifshitz symmetry \cite{Lifshitz;1941Fiz}, which is the symmetry under a transformation
\begin{align}
t\to \lambda^zt, \;\;x\to\lambda x,
\end{align}
where $z$ is the dynamical exponent, {and $\lambda$ is a positive scaling factor}. {For example, the Lifshitz symmetry can occur at some critical points in condensed matter systems \cite{PhysRevLett.35.1678}, and a quantum gravity model with the Lifshitz symmetry has been proposed in \cite{Horava:2009uw}.  Since Lorentz invariance is broken, the propagation speed of the quasiparticles and the behavior of the entanglement entropy in Lifshitz theories  may be different from the ones in Lorentz invariant theories. Thus, it is important to check such different behavior of the entanglement entropy with the Lifshitz symmetry.} The entanglement entropy in the Lifshitz theories was studied by {field-theoretical} methods and holographic methods in, for example, \cite{Solodukhin_2010, Keranen:2011xs, Fischetti:2014zja, Hosseini:2015gua, Gentle:2015cfp, Zhou:2016ykv, Kusuki:2017jxh, Gentle:2017ywk, MohammadiMozaffar:2017chk, Wen:2018mev}. The time-independent entanglement entropy in  the Lifshitz free scalar theories was studied in \cite{MohammadiMozaffar:2017nri, He:2017wla, Chen:2017txi, Chen:2017tij}, and the time dependent entanglement entropy in the sudden quench of the Lifshitz free scalar theories was studied in \cite{MohammadiMozaffar:2018vmk} with $z>1$ and in \cite{Nezhadhaghighi:2014pwa} with $0<z<1$.

In this paper, we study the time-dependent entanglement entropy on Lifshitz free scalar theories with $z>1$ in {1+1 dimensional} spacetime. {We compute the entanglement entropy by a correlator method, which is a computation method for free  theories}, on {1 dimensional spacial} lattice~\cite{Peschel_2009, Peschel:2003aa,  Casini:2009sr}. In order to obtain a perspective of continuum field theories from the computations on the lattice, we will take a smaller mass than the inverse lattice spacing to {suppress} cutoff effects. 

There is a related previous work on this topic {in \cite{MohammadiMozaffar:2018vmk}}, where only the sudden quench was considered. Here, we consider the slow and fast ECP and the slow and fast CCP. The sudden quench case in ~\cite{MohammadiMozaffar:2018vmk} can be obtained by the very fast limit of the ECP in our analysis. Another difference from ~\cite{MohammadiMozaffar:2018vmk} is the mass scales. While \cite{MohammadiMozaffar:2018vmk} deals with {the initial mass scale of order 1},  here we consider a small mass scale compared with the lattice spacing since we are interested in the {field} theory limit.

We have found many interesting features on the {dynamics} of the entanglement entropy: Some are independent of the subsystem size and some are independent of the dynamical exponent $z$. For example, at early times, in both ECP and CCP, the entanglement entropy is independent of the subsystem size and, at {late times}, in the CCP the entanglement entropy is oscillating in time with a well defined period, independently of $z$. We {will interpret} such properties by the quasiparticle picture and the idea of the correlator method.

In particular, there is an interesting distinctive property for $z>1$ compared with $z=1$ case. It is about a critical time $t_c$ that the entanglement entropy starts depending on the subsystem size significantly\footnote{{This `significantly' should be quantified properly. Here, we are more qualitative. It simply means it is observable from our numerical plot.}}. While $t_c$ for $z=1$ in the fast ECP and CCP can be explained well only by the fast quasiparticles of the quasiparticle picture, we find that this explanation does not work for $z>1$ and $t_c$ is delayed. We {will interpret} this by a careful investigation of the quasiparticle picture. Note that the similar delay has been observed in some spin chain models without the Lifshitz symmetry \cite{fagotti2008evolution, Alba:2017aa}.

The paper is organized as follows: In section \ref{sup} we review how to compute the entanglement entropy of the Lifshitz free scalar theories on 1 dimensional lattice by the correlator method. In section \ref{ECP} we compute the time evolution of the entanglement entropy for $z>1$ in the ECP and, in section \ref{CCP}, we do {so} in the CCP. In section \ref{sqpf} we study the quasiparticle formula in the sudden quench with $z=2$, by which we interpret our results in sections \ref{ECP} and \ref{CCP}.  We conclude in section \ref{summary}.

\section{Set up and method}\label{sup}
In this section, we consider a Hamiltonian of free scalar Lifshitz theories on 1 {dimensional} lattice. In order to study the time evolution {of the entanglement entropy}, we consider the end-critical-protocol (ECP) and cis-critical-protocol (CCP), where mass potentials  depend on time smoothly. We also define fast and slow limits of the ECP and CCP by using the relation between parameters in the mass potentials. Then, we explain how to compute the time evolution of entanglement entropy by using the correlator method\footnote{
In this paper we do not explain details of numerical calculations,
which can be found in Appendix D in the previous paper
\cite{Nishida:2017hqd}.
}.

\subsection{Hamiltonian and equation of Lifshitz free scalar theories}
In this subsection, we introduce a Hamiltonian of Lifshitz free scalar theories on $1$ {dimensional} {\it lattice} based on \cite{MohammadiMozaffar:2017nri, He:2017wla, MohammadiMozaffar:2018vmk}.
Let us first start with a Hamiltonian of Lifshitz free scalar field theories in 1 spacial dimension\footnote{Our convention is the same as one in \cite{He:2017wla}. {In this paper, we consider {$z \in \mathbb{Z}_{>0}$.}} }
\begin{equation}
\bar{H}(t)= \frac{1}{2}\int dx \left[\pi^2+\bar{\alpha}^2\left(\partial_x^z \phi\right)^2+\bar{m}(t)^{2}\phi^2\right],\label{H1}
\end{equation}
where {the overbar indicates dimensionful  observables, and $\bar{m}(t)$ is a mass potential which depends on $t$.} 

For numerical computations, we construct a Hamiltonian on $1$ {dimensional} lattice from (\ref{H1}). Let us consider $N $ lattice sites on a $1$ {dimensional} circle\footnote{
Here we impose the periodic boundary condition.
}
and discretize the system with {a} lattice spacing $\epsilon$. Accordingly,  by replacing $\int dx \to \epsilon \sum_{l=0}^{N-1}$, $\phi \to q_l$, {$\partial_x^z \phi \to \epsilon^{-z}\sum_{m=0}^z(-1)^{z+m}\binom{z}{m}q_{l+m-1}$}, $\pi \to p_l/\epsilon$, $\bar{\alpha} \rightarrow \alpha \epsilon^{z-1}$, $\bar{m}(t) \rightarrow m(t)/\epsilon$, and $\bar{H}(t) \rightarrow H(t)/\epsilon$, we obtain {a} lattice Hamiltonian on {a discretized} circle:
 \begin{align}
H(t)&=\frac{1}{2}\sum_{l=0}^{N-1}\left[p_l^2+\alpha^2\left(\sum_{m=0}^z(-1)^{z+m}\binom{z}{m}q_{l+m-1}\right)^2+m(t)^{2}q_l^2\right],\label{H2}
\end{align}
{where $\binom{z}{m}:=\frac{z!}{(z-m)!m!}$} {is the binomial coefficient}.
Note that, from here, all variables and parameters are dimensionless and dimensionful quantities are recovered by the lattice spacing $\epsilon$. 

To simplify the interaction between $q_l$ in the Hamiltonian (\ref{H2}), we use the Fourier transformations\footnote{We  assume that $N$ is an odd integer. {One can also do the similar analysis with even $N$.}}:
\begin{equation} \label{qandp}
\begin{split}
&q_l=\frac{1}{\sqrt{N}}\sum_{\k=-\frac{N-1}{2}}^{\frac{N-1}{2}}e^{i \frac{2\pi l \k}{N}}\tilde{q}_{\k}\,, \\
&p_l=\frac{1}{\sqrt{N}}\sum_{\k=-\frac{N-1}{2}}^{\frac{N-1}{2}}e^{i \frac{2\pi l \k}{N}}\tilde{p}_{\k}\,,
\end{split}
\end{equation}
and we obtain
\begin{align}
H(t)=\frac{1}{2}\sum_{\k=-\frac{N-1}{2}}^{\frac{N-1}{2}}\left[\tilde{p}_{\k}^{\dagger}  \tilde{p}_{\k}+ \left(\alpha^2\left(2\sin{\left(\frac{\pi {\k}}{N}\right)}\right)^{2z}+ m^2(t)\right)\tilde{q}_{\k}^{\dagger}\tilde{q}_{\k}\right].\label{H59}
\end{align}
{Throughout this paper, we take $\alpha=1$ without loss of generality because results for  other values of $\alpha$ can be obtained by  rescaling $m(t)$ and time.}

We expand $\tilde{q}_{\k}$ and $\tilde{p}_{\k}$ by a creation operator $a^{\dagger}_{k}$ and an annihilation operator $a_{k}$ as
\begin{equation} \label{eqqp}
\begin{split}
&\tilde{q}_{\k} =f_{\k}(t) a_{\k} + f_{-\k}^*(t) a^{\dagger}_{-\k}\,, \\
&\tilde{p}_{\k}=\dot{f}_{\k}(t) a_{\k} + \dot{f}_{-\k}^*(t) a^{\dagger}_{-\k}\,,
\end{split}
\end{equation}
and we quantize them by the canonical commutation relations $[\tilde{q}_\a, \tilde{p}_\b]= i \delta_{\a, -\b}$, $[a_{\a}, a^{\dagger}_{\b}]=\delta_{\a, \b},$ and  $[\tilde{q}_\a, \tilde{q}_\b]= [\tilde{p}_\a, \tilde{p}_\b]=[a_{\a}, a_{\b}]= [a^{\dagger}_{\a}, a^{\dagger}_{\b}]=0$. 
From the Heisenberg equations of \eqref{eqqp} with (\ref{H59}), the equation of $f_k(t)$ yields
\begin{equation} \label{eofk}
\begin{split}
&\frac{d^2 f_{k}(t)}{dt^2} + \omega_k^2(t) f_{k}(t)=0\,,\\
&\omega_k(t)=\sqrt{\left(2\sin{\left(\frac{k}{2}\right)}\right)^{2z}+ m^2(t)}\,.
\end{split}
\end{equation}
{Here, we introduce} the rescaled momentum {$k$ as}
\begin{equation}
k := \frac{2 \pi \k}{N}  \,.
\end{equation}

\subsection{Two mass potentials: ECP and CCP}
For numerical computations, we use smooth\footnote{Contrary to `smooth', we may consider the `sudden' quench, which is realized by a step function.} mass potentials in which $f_k(t)$ has analytic solutions of (\ref{eofk}). One of them is the mass potential in the end-critical-protocol (ECP) \cite{Chandran_2012}:
\begin{align}
m^2(t)=\frac{m_0^2}{2}\left[1-\tanh{\left(\frac{t}{\delta t}\right)}\right].\label{mecp}
 \end{align}
In the ECP, the initial mass is $m_0$, and the mass potential decreases with time and becomes zero at late times as shown in the left panel of Figure \ref{quench}. Another mass potential with which we can obtain an analytic solution of (\ref{eofk}) is the mass potential in the cis-critical-protocol (CCP) \cite{Chandran_2012}:
\begin{align}
m^2(t)=m^2_0\tanh^2{\left(\frac{t}{\delta t}\right)}.\label{mccp}
\end{align}
In the CCP, the initial and final masses are $m_0$, and the mass potential at $t=0$ becomes zero as shown in the right panel of Figure \ref{quench}. 


An explicit solution of (\ref{eofk}) in the ECP is \cite{Bernard:1977pq},
\begin{equation}\label{fkECP}
\begin{split}
 f_k(t)=&\frac{1}{\sqrt{-4i\beta/\delta t}}\biggl( \frac{1+{\rm tanh}[t/\delta t]}{2}\biggr)^{-\beta} \biggl( \frac{1-{\rm tanh}[t/\delta t]}{2}\biggr)^{-\alpha} \\
 &\times~_2 F_1(-\alpha - \beta +1 , -\alpha -\beta; -2\beta+1;(1+{\rm tanh}[t/\delta t])/2),\\
 \alpha:=& -\frac{i\delta t}{2}|2{\rm sin}[k/2]|^z,~\beta:=\frac{i\delta t}{2}\sqrt{ (2{\rm sin}[k/2])^{2z} + m_0^2  },
\end{split}
\end{equation}
and one in the CCP is \cite{Das:2014hqa},
\begin{equation} \label{fkCCP}
\begin{split} 
f_{k}(t) =&\frac{2^{{\rm i} \omega_0  \delta t}}{\sqrt{2\omega_{0}}}\frac{({\rm cosh}[t/\delta t])^{2\alpha}}{E_{1/2}E'_{3/2}-E_{3/2}E'_{1/2}} \times\biggl[E'_{3/2}~ _2 F_1 \biggl(a,b;\frac{1}{2} ;-{\rm sinh}^2[t/\delta t] \biggr) \\
&+E'_{1/2}{\rm sinh}[t/\delta t] _2 F_1 \biggl(a+\frac{1}{2},b+\frac{1}{2};\frac{3}{2} ;-{\rm sinh}^2[t/\delta t] \biggr)
\biggl],\\
&a := \alpha -\frac{ i \omega_0 \delta t}{2}, \;\;b := \alpha +\frac{ i\omega_0 \delta t}{2},  \\
& \alpha := \frac{1+\sqrt{1-4(m_0\delta t)^2}}{4},\;\;\omega^2_0 := (2{\rm sin}[k/2])^{2z} +m_0^2 ,  \\
&E_{1/2}:=\frac{\Gamma(1/2)\Gamma(b-a)}{\Gamma(b)\Gamma(1/2-a)}
, \;\;E_{3/2}:=\frac{\Gamma(3/2)\Gamma(b-a)}{\Gamma(1/2+b)\Gamma(1-a)}
, \;\;E'_{c}:=E_c (a\leftrightarrow b). 
\end{split}
\end{equation}

\subsection{Fast and slow limits}

The mass potentials (\ref{mecp}) and (\ref{mccp}) depend on $m_0$ and $\delta t$, where we define an initial ($t \rightarrow -\infty$) length scale $\xi$ as $\xi:=1/m_0$.  By using these parameters, we define two limits of the quenches as in \cite{Das:2016lla}: fast and slow limits. The fast limit is defined such that $\delta t$ is much smaller than the initial length scale $\xi$, \textit{i.e.},
\begin{align}
\delta t\ll\xi \,,
\end{align}
while the slow limit is defined such that $\delta t$ is much larger than the initial length scale $\xi$ as
\begin{align}
\delta t\gg\xi \,.
\end{align}

One characteristic difference between the fast and slow limit is time scales when the adiabaticity breaks. {To define the time scale,} we use a {dimensionless} function
\begin{align} \label{lcr}
C_{\rm{L}}(t):=\left|\frac{1}{m^2(t)}\times \frac{d m(t)}{dt}\right|
\end{align} 
for a criteria of the adiabaticity (Landau criteria). {See \cite{Gritsev:2009wt, Das:2016lla, Fujita:2018lfj}, {for details.}}   If $C_{\rm{L}}(t)$ satisfies $C_{\rm{L}}(t)\ll1$, we can use the adiabatic expansion {because} the adiabaticity is {held}. The Kibble-Zurek time $t_{\rm{kz}}$\footnote{{This $t_{\rm{kz}}$ is determined from $\omega_k(t)$ at  $k=0$. See section 3 in  \cite{Fujita:2018lfj} for more detail.}} is defined such that
\begin{align}
C_{\rm{L}}(t_{\rm{kz}})\sim1\,,
\end{align}
which means that $t_{\rm{kz}}$ is the time scale when the adiabaticity starts breaking (or being restored in the case of CCP).  

In the fast limit $t_{\rm{kz}} \sim 0$, {while $t_{\rm{kz}}$ in the slow limit} is far from $t=0$.  The Kibble-Zurek time  $t_{\rm{kz}}$ in the slow ECP and CCP is \cite{Das:2016lla}
\begin{align}
&t_{\text{kz}}\sim \delta t \log[\delta t/\xi] \;\;(\textrm{ECP}),
\label{ECPkz}\\ 
&t_{\text{kz}}\sim (\delta t\xi)^{\frac{1}{2}} \;\;(\textrm{CCP}).
\label{CCPkz}
\end{align}
In the slow ECP, the adiabaticity is broken after $t\sim t_{\rm{kz}}$, and the one in the slow CCP is broken from $t\sim -t_{\rm{kz}}$ to $t\sim t_{\rm{kz}}$.
For later use, we here define a length scale $\xi_{\rm{kz}}$ at $t=t_{\rm{kz}}$ as
\begin{align}
\xi_{\text{kz}}:=&\frac{1}{m(t_{\text{kz}})}\sim\delta t\;\;(\textrm{ECP}),\\
\xi_{\text{kz}}:=&\frac{1}{m(t_{\text{kz}})}\sim(\delta t\xi)^{\frac{1}{2}},\;\;(\textrm{CCP}).
\end{align}

\subsection{Correlator method}
In free scalar theories, we can compute the entanglement entropy by using {two-point} functions. This computation method is called {as} the correlator method \cite{Peschel_2009, Peschel:2003aa,  Casini:2009sr}, and here we review this method based on \cite{Coser:2014gsa}.

In our computations, we consider a thermodynamic limit $N\to\infty$ with  fixed $\epsilon$. In this limit, \eqref{qandp} is written as
\begin{align}
q_l(t) :=\int^{\pi}_{-\pi} \frac{d{k}}{\sqrt{2\pi}}\tilde{q}_{k} e^{i {k} l}\,, \\
p_l(t) := \int^{\pi}_{-\pi} \frac{d{k}}{\sqrt{2\pi}}\tilde{p}_{k} e^{i {k} l}\,,
\end{align}
and {two-point} functions of $q_l(t)$ and $p_l(t)$ are
\begin{align}\label{spect}
&Q_{ab}(t):=\left\langle0| q_a (t)q_b(t) |0\right \rangle  =\int^{\pi}_{-\pi} \frac{dk}{2\pi}\left|f_{k}(t)\right|^2\cos{\left(k\left|a-b\right|\right)}, \\
&P_{ab}(t):=\left\langle0| p_a (t)p_b(t) |0\right \rangle = \int^{\pi}_{-\pi} \frac{dk}{2\pi}\left|\dot{f}_{k}(t)\right|^2\cos{\left(k\left|a-b\right|\right)}, \\
&D_{ab}(t):=\frac{1}{2}\left\langle0| \left\{q_a (t), p_b(t) \right\} |0\right \rangle =\int^{\pi}_{-\pi} \frac{dk}{2\pi} \mathrm{Re}\left[\dot{f}^*_{k}(t)f_{k}(t)\right]\cos{\left(k\left|a-b\right|\right)}, 
\end{align}
where $|0\rangle$ is {the ground state for the initial Hamiltonian}. With the explicit expressions of $f_k(t)$ in the ECP (\ref{fkECP}) and the CCP (\ref{fkCCP}), these {two-point} functions can be computed numerically.

In the correlator method, the entanglement entropy of the subsystem $A$ with the number of lattice sites $l$ can be computed by the eigenvalues of a matrix $\mathcal{M}$ constructed from the two point functions,
\begin{align}
\mathcal{M}:=iJ\Gamma\,, \quad &J:=\begin{bmatrix}
0 & I_{l \times l} \\
-I_{l \times l} & 0 \\
\end{bmatrix}\,, \quad 
\Gamma:=\begin{bmatrix}
Q_{ab}(t) & D_{ab}(t) \\
D_{ab}(t) & P_{ab}(t) \\
\end{bmatrix},
\end{align}
 where $I_{l \times l}$ is an $l\times l$ unit matrix. 
By computing positive eigenvalues of the $2l\times2l$ matrix $\mathcal{M}$, say $\gamma_a$, we {obtain} 
the entanglement entropy $S_A(t)$ for a subsystem $A$ as follows:
\begin{align}
S_A(t)=\sum_{a=1}^{l}\left[\left(\gamma_{a}+\frac{1}{2}\right)\log{\left(\gamma_{a}+\frac{1}{2}\right)}-\left(\gamma_{a}-\frac{1}{2}\right)\log{\left(\gamma_{a}-\frac{1}{2}\right)}\right]\,. \label{sAcm}
\end{align}
 
In this paper, we study the time evolution of the entanglement entropy in the free Lifshitz scalar theories. In order to study the time evolution, we  compute a change of {the} entanglement entropy 
{
\begin{equation}
\Delta S_A(t)=S_A(t)-S_A(-\infty).
\end{equation}}
In particular, we investigate  {$z$-dependence and $l$-dependence of $\Delta S_A$. }

\section{Entanglement entropy  in the ECP with $z>1$}\label{ECP}
In this section, we first describe our numerical results on the time evolution of the entanglement entropy in the fast and slow ECP. {After that}, we provide interpretations of our results. 

\subsection{Fast ECP} \label{FECP1}
As an example of the fast ($\delta t /\xi \ll 1$) ECP  with $z \ne 1$ (Lifshitz theory){,} we choose 
\begin{equation}
\xi=100\,, \quad \delta t=5\,.
\end{equation}

Figure \ref{ecpf1} shows  {$l$-dependence} of $\Delta S_A$ for $z=2$ with {$l=100, 200, 300, 400, 500, 1000, 2000$.}   The entanglement entropy in the fast ECP with small $\delta t/\xi$ is similar to one in the sudden quench because the sudden quench {is expected to be  a limit} of the ECP with $\delta t\to0$.  
We observe the following properties {from} Figure \ref{ecpf1}:

\begin{figure}[]
 \centering
     \subfigure[{Comparison between different subsystem sizes $l$.}]
     {{\includegraphics[width=7.3 cm]{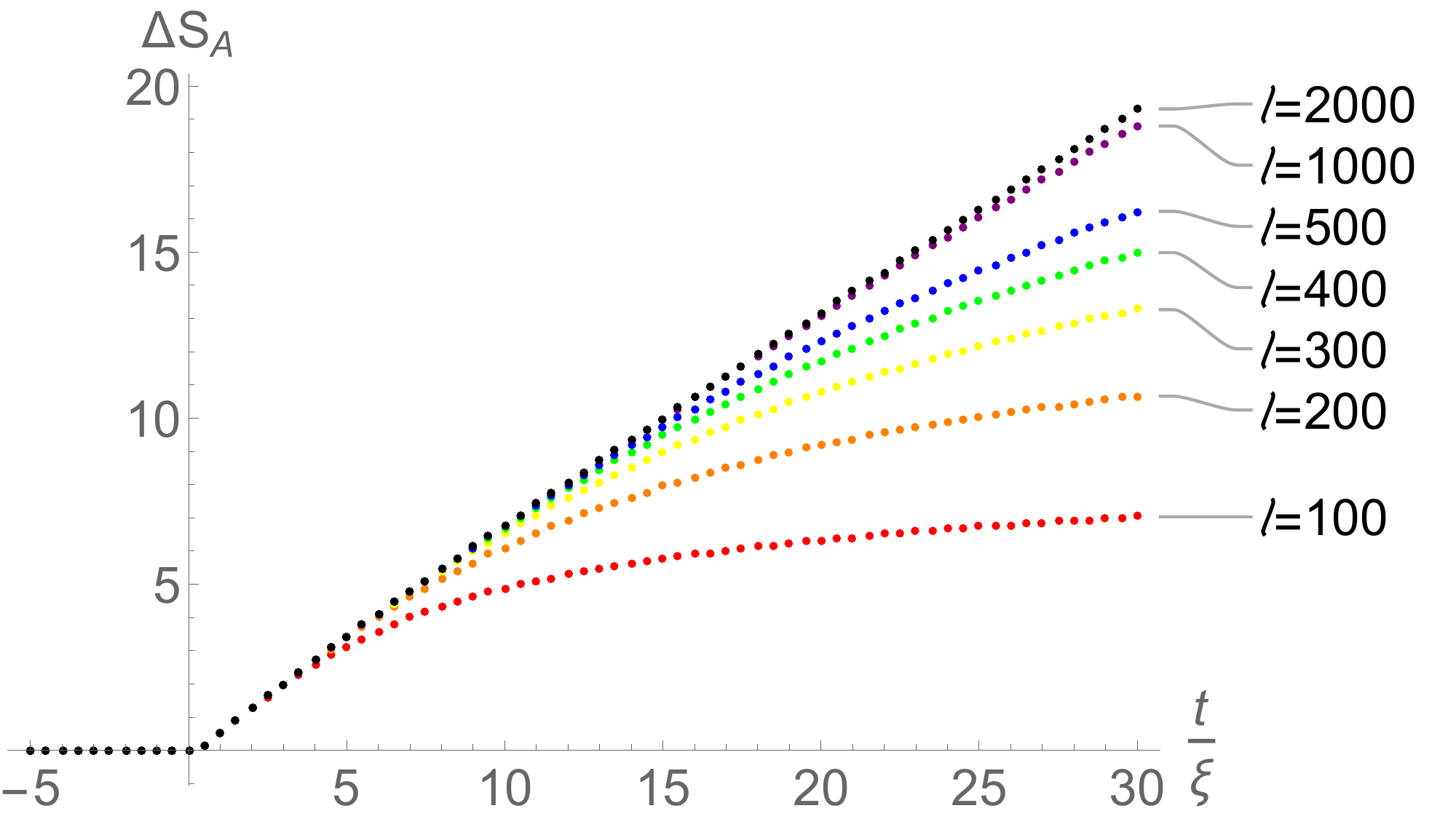} \label{fig2a}} }
     \subfigure[{Comparison between $l=1000, 2000$. The dashed line designates the critical time for $z=1$: $\frac{t_c}{\xi}\sim\frac{l}{2\xi} = 5$ for $l=1000$.}]
     {{\includegraphics[width=7.3 cm]{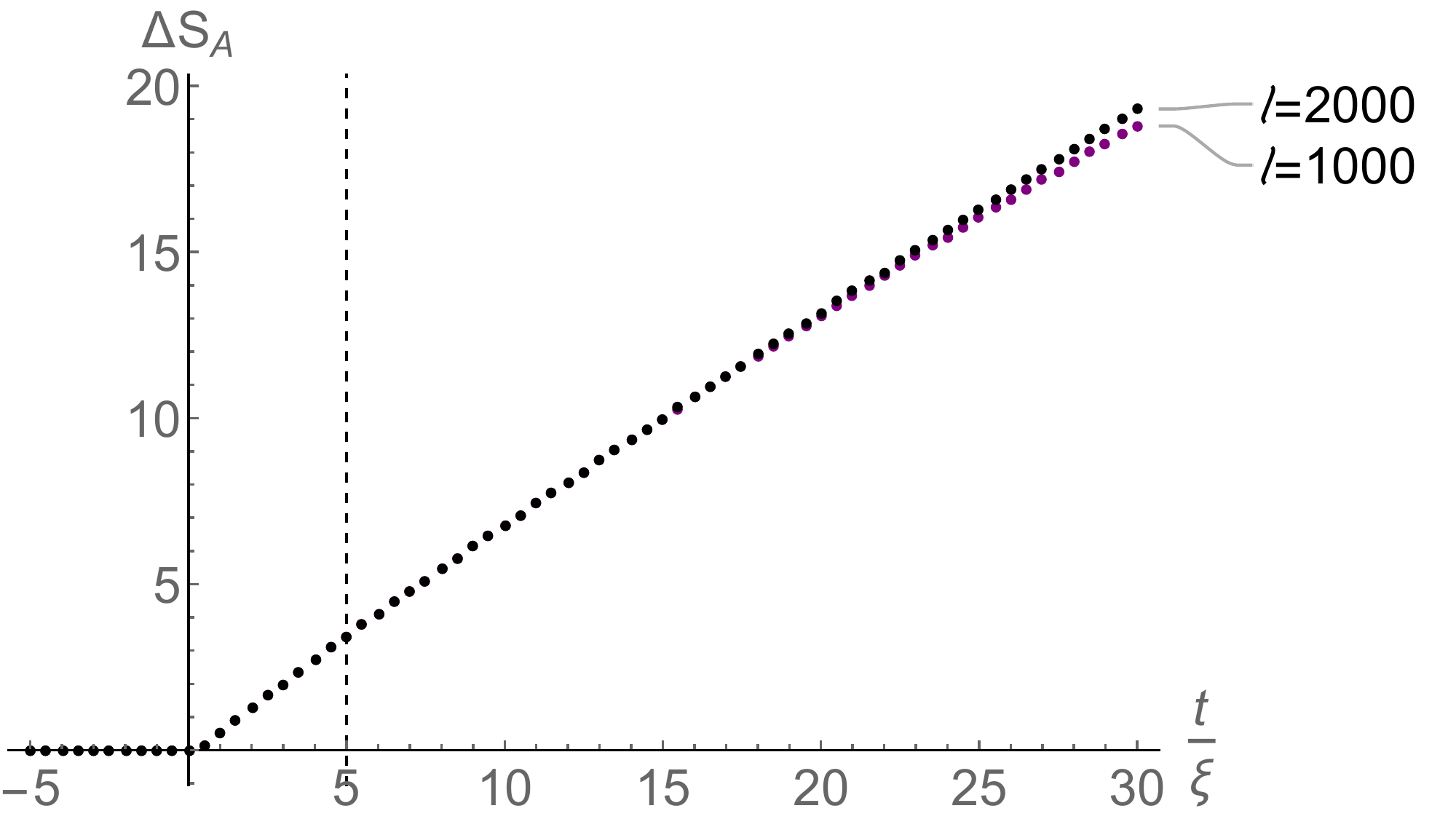} \label{fig2b}}}
           \caption{{Time dependence of $\Delta S_A$ in the fast ECP for $z=2$ ($\xi=100, \delta t=5$).} }
  \label{ecpf1}
\end{figure}
\begin{figure}[]
 \centering
     {{\includegraphics[width=7.3 cm]{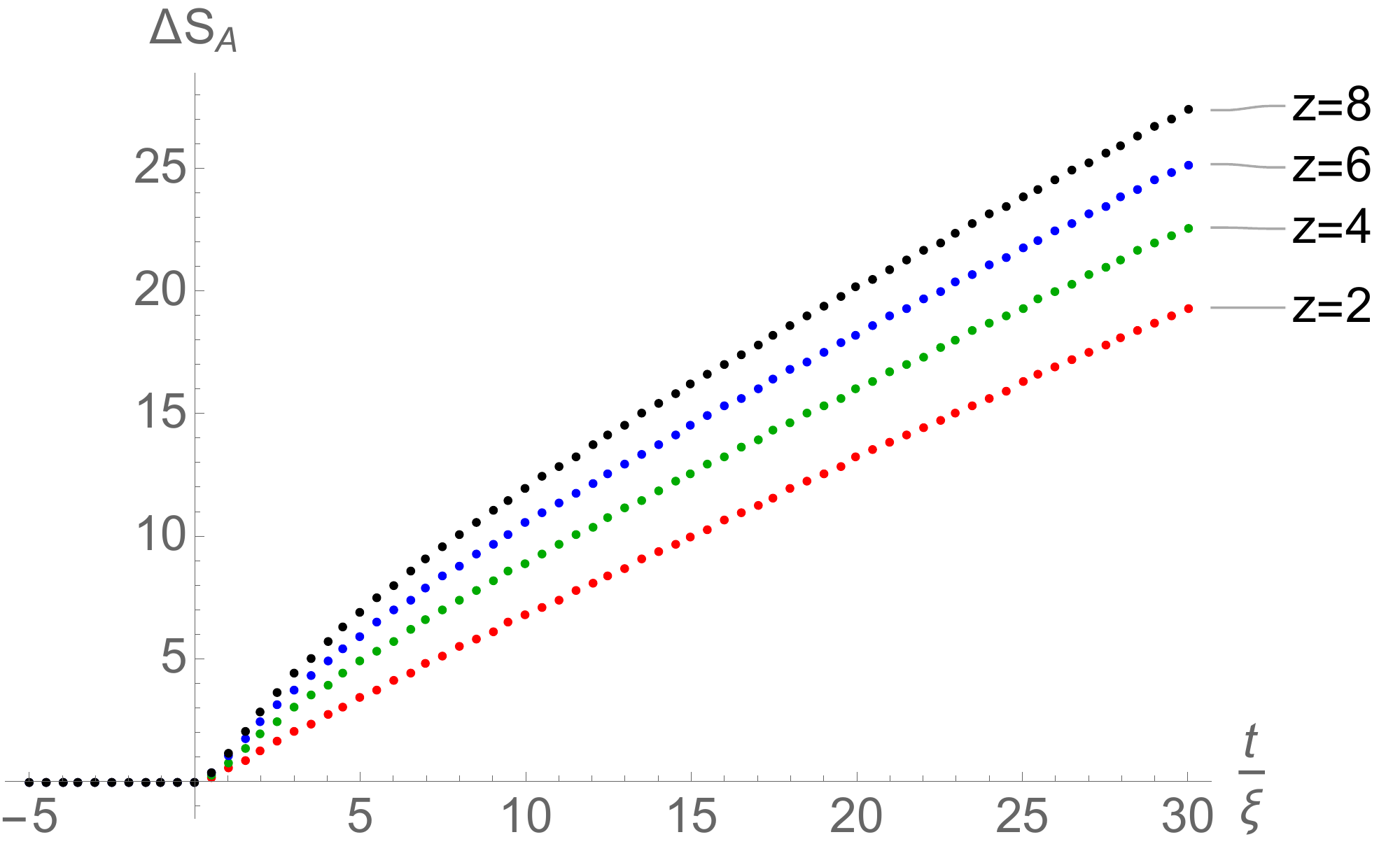} \label{}}}
           \caption{{Time dependence of $\Delta S_A$ in the fast ECP with $\xi=100, \delta t=5, l=2000$ and $z=2,4,6,8$.}}
  \label{ecpf2}
\end{figure}

\begin{enumerate}
\item[(Ef1)] {The change of the entanglement entropy} $\Delta S_A$ begins to increase around $t\sim0$ like the sudden quench case. 
\item[(Ef2)] At early times, all plots lie on the same curve independently of the subsystem size $l$. It means $\Delta S_A$ has no subsystem size-dependence at early times.
\item[(Ef3)] At late times, $\Delta S_A$ with the different subsystem sizes {is} different. {The critical time $t_c$, when the significant subsystem size-dependence of $\Delta S_A$ occurs,} increases with the subsystem size. 
For $z=1$, it is expected that $t_c(z=1)\sim l/2$ from the quasiparticle picture~\cite{Nishida:2017hqd}. For $z=2$, we find  {$t_c(z=2) > t_c(z=1)$}. For example, see Figure \ref{fig2b}, where $t_c(z=1)/\xi \sim 5${, while} $t_c(z=2)/\xi \sim 25$ for $l=1000$.
\end{enumerate}

Figure \ref{ecpf2} shows  {$z$-dependence} of $\Delta S_A$ for $l=2000$ with $z=2,4,6,8$. The change 
$\Delta S_A$ in this figure has the following properties of the $z$-dependence:

\begin{enumerate}%
\item[(Ef4)] As {$z$} increases, $\Delta S_A$ also increases.
\item[(Ef5)] 
In the figure, we focus on the time range before {$l$-dependence}  appears significantly. At late times, $\Delta S_A$ linearly increases with $t$, while  $\Delta S_A$ at early times increases nonlinearly.  {This nonlinearity  is sustained for {a wide time range as  $z$} increases.}
\end{enumerate}

\subsection{Slow ECP}
As an example of the slow ($\delta t /\xi \gg 1$) ECP  with $z \ne 1$ (Lifshitz theory){,} we choose 
\begin{equation}
\xi=5\,, \quad \delta t=500\,.
\end{equation}

Figure \ref{ecps1} shows  {$l$-dependence} of $\Delta S_A$ for $z=2$ with $l=10,50,$ $100,300,1000,2000$.  
We observe the following properties {from} Figure \ref{ecps1}:

\begin{figure}[]
 \centering
     \subfigure[{Comparison between different subsystem sizes $l$.}]
     {{\includegraphics[width=7.3 cm]{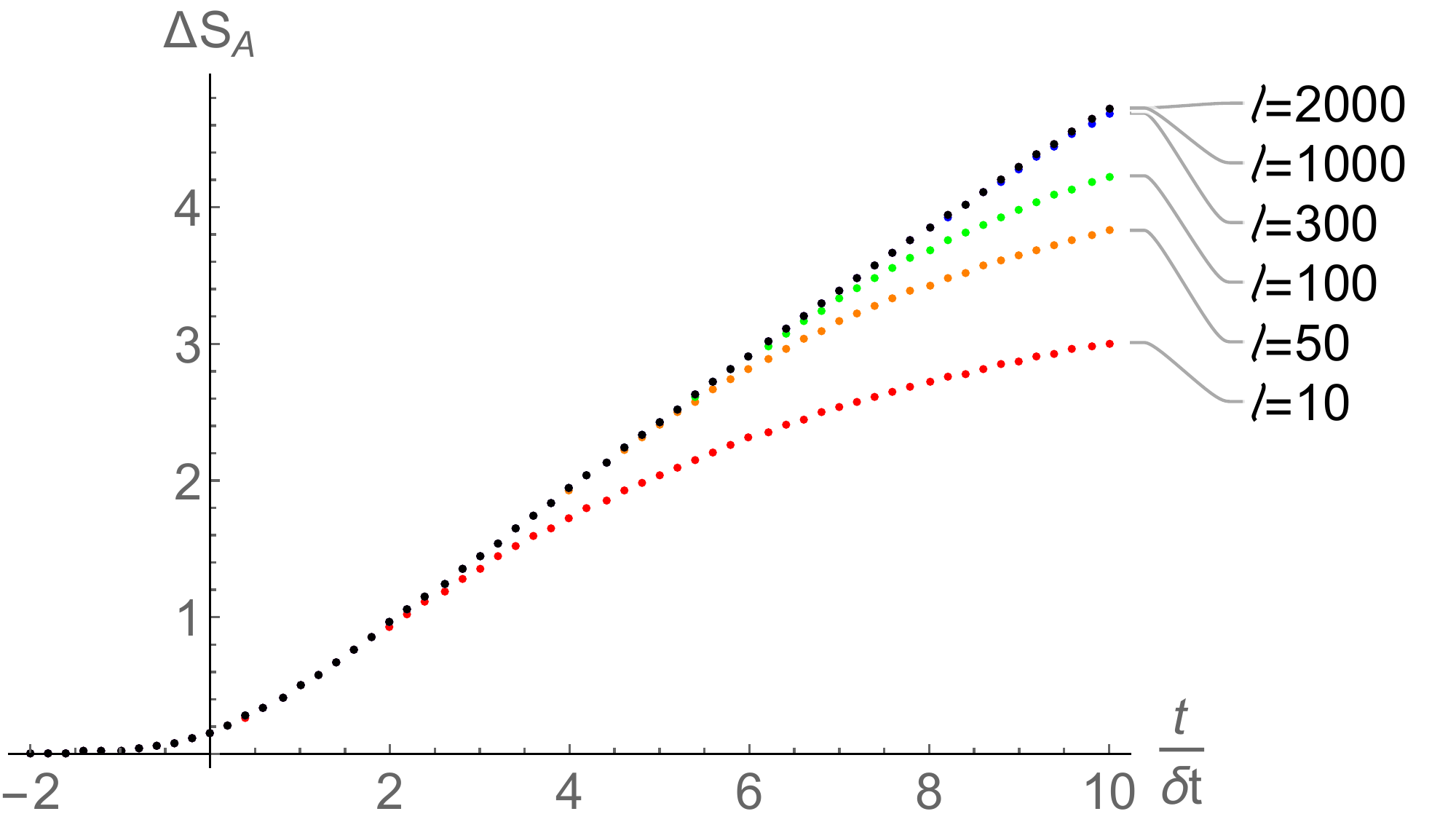} \label{}} }
     \subfigure[{Comparison between $l=1000, 2000$. The dashed line designates the critical time for $z=1$: $\frac{t_c}{\delta t}\sim \frac{t_{\textrm{kz}}}{\delta t} +  \frac{l}{2\delta t} \sim 4.8$ for $l=1000$ ($t_{\textrm{kz}}\sim 2300$).} ]
     {{\includegraphics[width=7.3 cm]{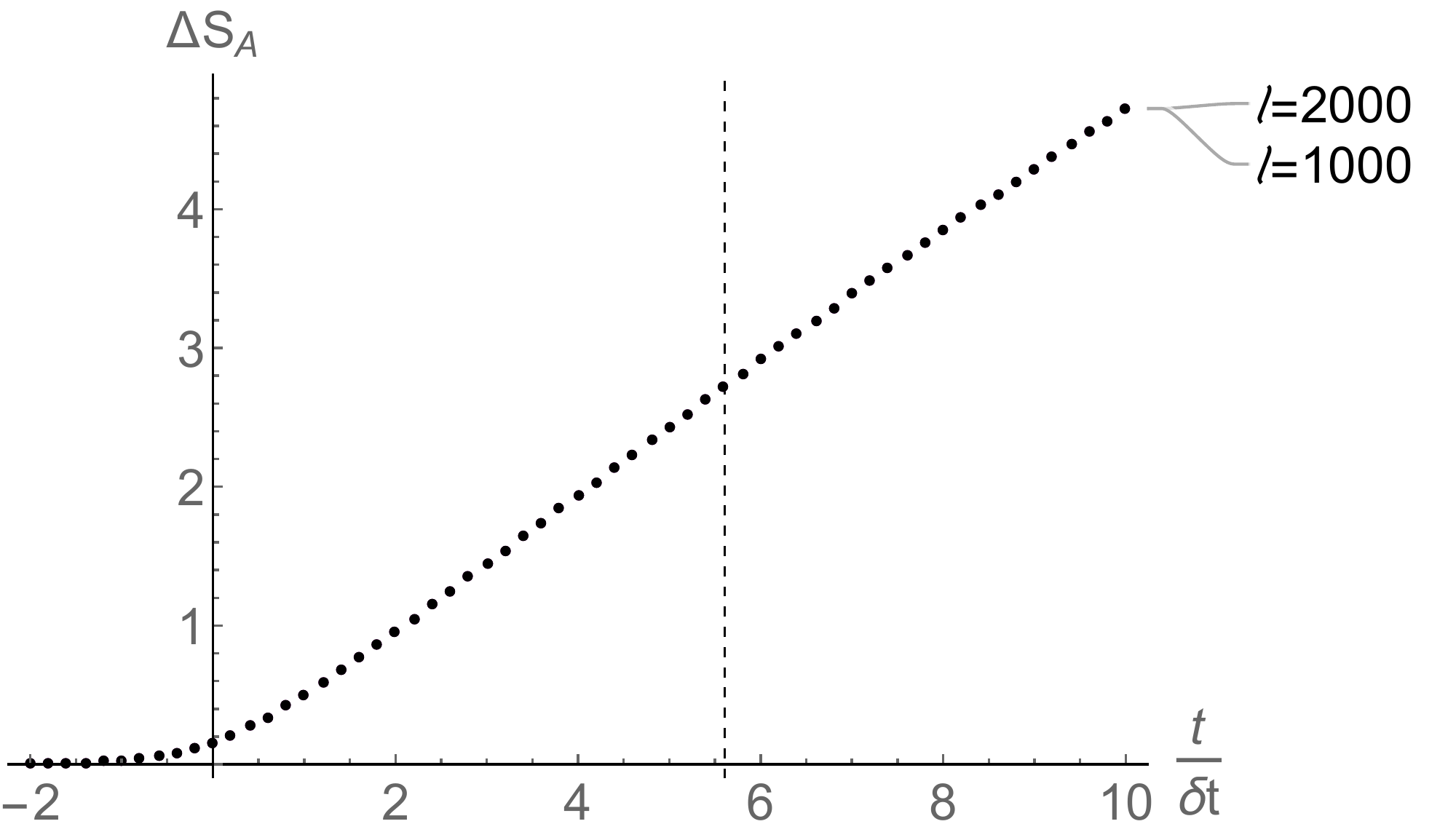} \label{fig4b}}}
           \caption{{Time dependence of $\Delta S_A$ in the slow ECP for $z=2$ ($\xi=5, \delta t=500$).}  }
  \label{ecps1}
\end{figure}
\begin{figure}[]
 \centering
     {{\includegraphics[width=7.3 cm]{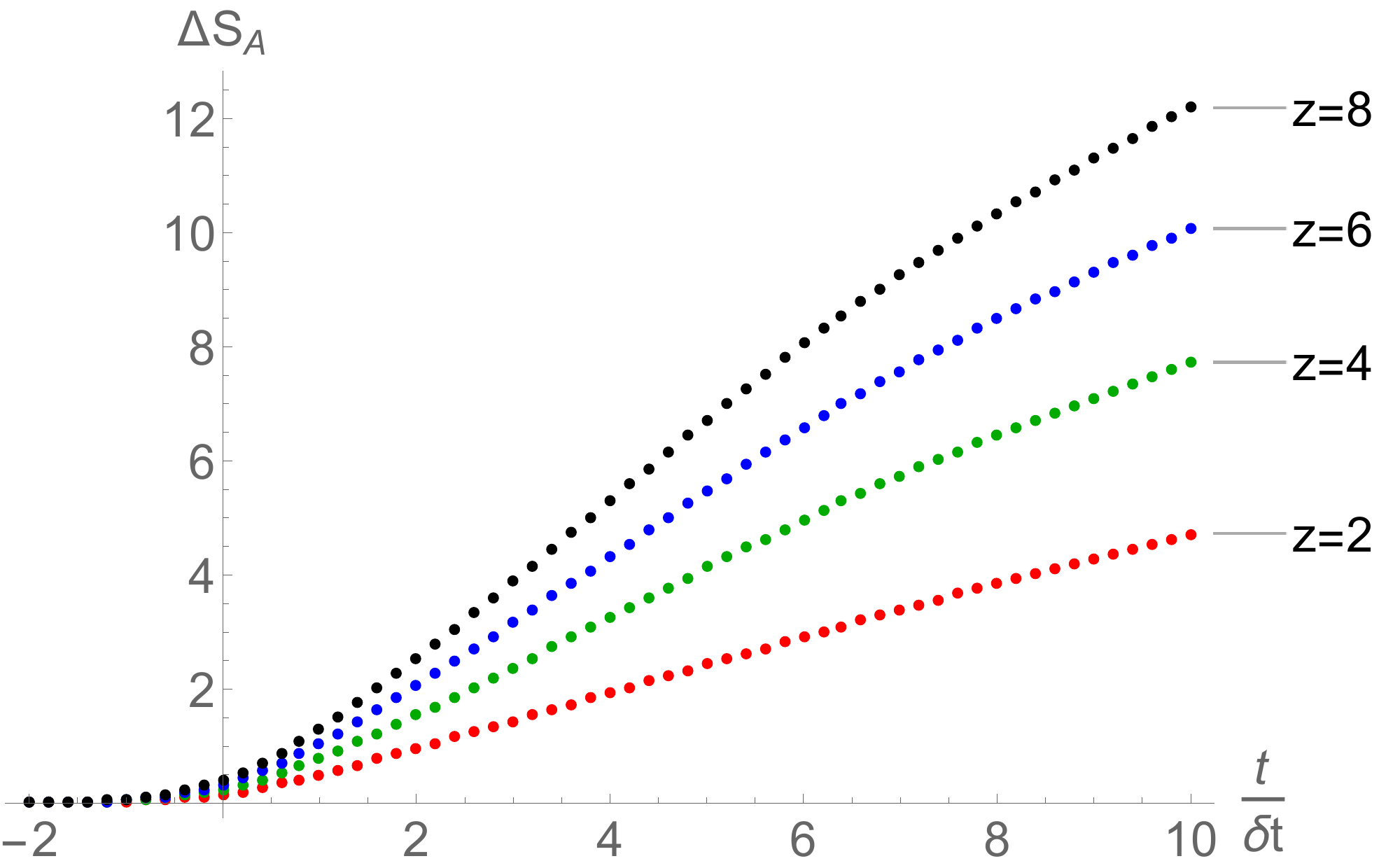} \label{}}}
           \caption{{Time dependence of $\Delta S_A$ in the slow ECP with $\xi=5, \delta t=500, l=1000$ and $z=2,4,6,8$.}}
  \label{ecps2}
\end{figure}

\begin{enumerate}
\item[(Es1)] {The change} $\Delta S_A$ in the slow ECP starts increasing at $t<0$, unlike the fast or sudden quench cases. 
\item[(Es2)] At early times $\Delta S_A$ has no subsystem size-dependence, like the fast ECP.
\item[(Es3)] At late times, $\Delta S_A$ with the different subsystem sizes are different and {the critical time $t_c$} increases with the subsystem size, like the fast ECP.
{For $z=1$ and large $l$}, it is expected that $t_c(z=1)\sim t_{\textrm{kz}} + l/2$ from the quasiparticle picture~\cite{Nishida:2017hqd}. For $z=2$, we find that $t_c(z=2) > t_c(z=1)$. For example, see Figure \ref{fig4b}, where $t_c({z=1})/\delta t \sim 4.8$ ($t_{\textrm{kz}}\sim 2300$ by \eqref{ECPkz} and $l=1000$) while $t_c(z=2)/\delta t > 10$.
\end{enumerate}

Figure \ref{ecps2} shows {$z$-dependence} of $\Delta S_A$ for $l=1000$ with $z=2,4,6,8$.
{The change} $\Delta S_A$ in this figure has the following properties of the $z$-dependence:

\begin{enumerate}
\item[(Es4)] Like the fast ECP,  $\Delta S_A$ increases as  {$z$} becomes large.
\item[(Es5)] 
In the figure, we focus on the time range before {$l$-dependence} appears significantly. {Like the fast ECP, $\Delta S_A$ increases nonlinearly, and this nonlinearity is sustained for a wide time range  as {$z$} increases.}
\end{enumerate}

\subsection{Interpretation of the properties}\label{IECP}
In this subsection, we interpret the aforementioned properties of the entanglement entropy for the ECP.

\paragraph{(Ef1) and (Es1)} The mass potential $m^2(t)$ starts decreasing around $t\sim -\delta t$. Therefore, $\Delta S_A$ in the slow ECP starts increasing {at} $t<0$.  For the fast ECP, however, since $\delta t \sim 0$, $m^2(t)$ {starts} decreasing around $t\sim 0$, {and} $\Delta S_A$ starts increasing around $t\sim 0$. 

\paragraph{(Ef2) and (Es2)}
To understand the subsystem size-independence of $\Delta S_A$ at early times we use the quasiparticle picture \cite{Calabrese:2005in, Alba:2017aa, Alba:2017lvc}. 
%
To explain the quasiparticle picture, let us consider a one-dimensional system with a subsystem $A$ of length $l$ under a {sudden} quench, where the mass potential changes at $t=0$ from the initial mass $m_0$ to the final mass $m_f=0$.
Due to a sudden quench at $t=0$, {the quasiparticle pairs} are created at $t=0$ everywhere and then propagate with the group velocity $v_k=\frac{d \omega_k}{d k}$ with the momentum $k$ which is computed by the Hamiltonian after a quench. 
\begin{figure}[]
 \centering
     {{\includegraphics[width=12 cm]{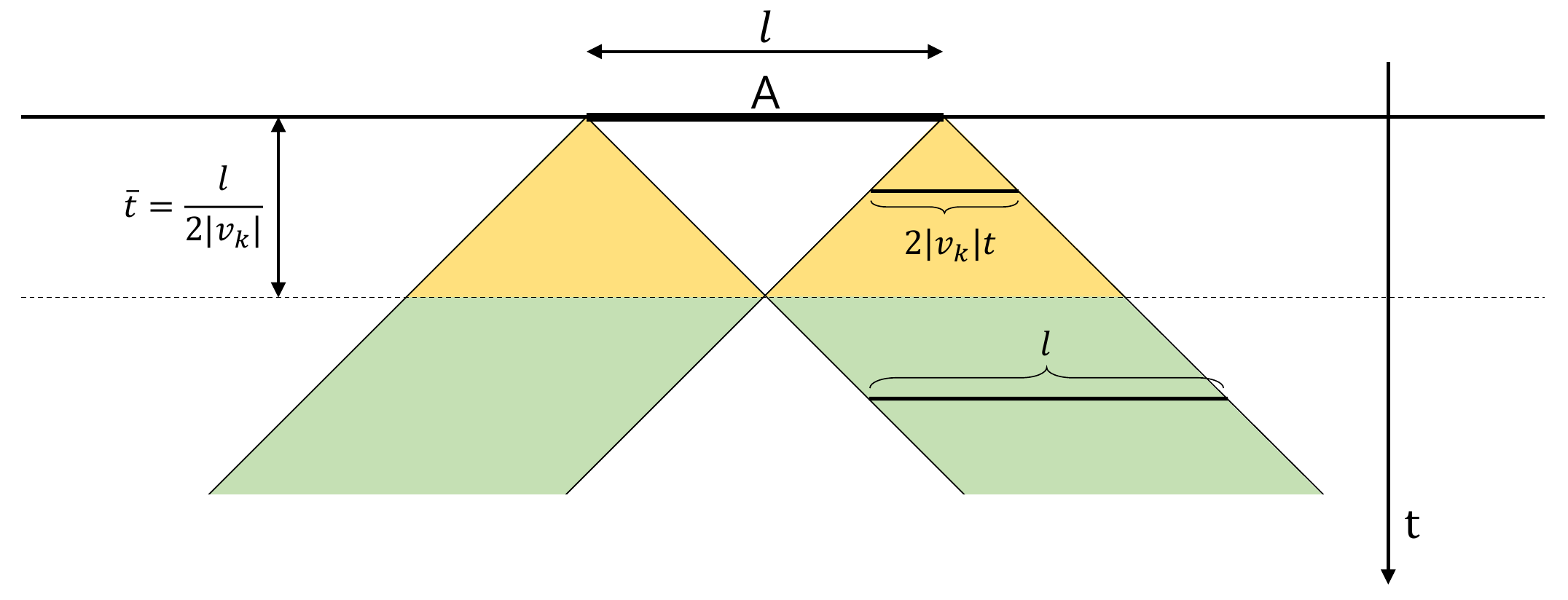} \label{}}}
           \caption{{Quasiparticle picture in the sudden quench}. The length of the subsystem $A$ is $l${, and} $v_k$ is the group velocity of the quasiparticle pairs created at $t=0$.  {In $t < \bar{t} $ the range of $2 |v_k|t$ (yellow area) contributes to the entanglement entropy, while in $t > \bar{t} $ the range of $l$ (green area) contributes to the entanglement entropy}.}
  \label{quasip}
\end{figure}

As shown in Figure \ref{quasip}, at the time $t$, only the quasiparticle pairs created in the yellow and green {region} will contribute to the entanglement entropy because one of the pairs is inside the subsystem $A$ and the other is outside of $A$. In other words,  {the quasiparticle pairs created in the length of $2|v_k|t$ contribute {in early time regime} $t< \bar{t} := \frac{l}{2|v_k|}$ (yellow area in Figure  \ref{quasip}), while  the quasiparticle pairs created in the length of $l$ contribute in late time regime $t>\bar{t}$ (green area).} 
Therefore, the  entanglement entropy for the subsystem $A$,  {{\it generated by the quasiparticle pairs with the group velocity $v_k$}}, depends on the subsystem size $l$ of $A$ after $t=\bar{t} = \frac{l}{2 |v_k|}$ and does not depend on the subsystem size in early time regime $t < \bar{t}$. 
It {turns} out that this quasiparticle picture {is} consistent with the entanglement entropy  with $z=1$ in the 2 dimensional CFT \cite{Calabrese:2005in}, the sudden quench \cite{Nezhadhaghighi:2014pwa}, and the ECP \cite{Nishida:2017hqd}.

\paragraph{(Ef3)} {In general, the group velocity $v_k$ depends on the momentum $k$ of the quasiparticle, {thus} we need to consider the quasiparticles with various velocities.  In order to determine {the critical time $t_c$}, we may use the maximum group velocity $v_{\textrm{max}}$,}{
\begin{equation} \label{tt7}
t_c \sim  \textrm{min}\{\bar{t}\} = \frac{l}{2 |v_\textrm{max}|} \,,
\end{equation}
}{
because it is the earlist time the subsystem size dependence enters.

For example, since $v_{\textrm{max}} =1$ for the massless quasiparticle with $z=1$,  $t_c$ with $z=1$ becomes $t_c\sim\frac{l}{2}$ from (\ref{tt7}), which is shown in Figure \ref{fig2b}.
%
%
{For $z>1$, the maximum group velocity is $|v_{\textrm{max}}| > 1$, so we expect $t_c $ to obey $t_c < l/2$.} However,   Figure \ref{fig2a} shows  
\begin{equation} \label{z200}
t_c > \frac{l}{2}  \qquad (z=2)\,.
\end{equation}
{We also confirmed this delayed $t_c$ for $z=4, 6, 8$.}

One possible {interpretation} of this delayed $t_c$ in the quasiparticle picture is as follows. The entanglement entropy by the quasiparticle pairs {comes from not only the ones with the fast velocity close to $|v_{\textrm{max}}|$, but also the sum of all quasiparticles with various $v_k$.} If the contribution of the fast quasiparticles to $\Delta S_A$  can be suppressed compared with the slow quasiparticles, the subsystem size-dependence around $t_c\sim \frac{l}{2 |v_\textrm{max}|}$ for the fast ECP may be negligible. Thus, if this conjecture works, in more general, we {expect $t_c$ to satisfy
{
\begin{equation} \label{z201}
  t_c \gg  \frac{l}{2 |v_\textrm{max}|} \,,
\end{equation}
}where $t_c \gg  \frac{l}{2 |v_\textrm{max}|}$ means that $t_c$ is large enough so that  we can observe difference between $t_c$ and $\frac{l}{2 |v_\textrm{max}|}$ from numerical plots.}
Note that the condition \eqref{z200} may not be satisfied even if \eqref{z201} is satisfied because it is still possible, in principle, 
\begin{equation} \label{z202}
\frac{l}{2 |v_\textrm{max}|}  < t_c  < \frac{l}{2} \,,
\end{equation}
for $|v_\textrm{max}| > 1$. 
In section \ref{sqpf}, we show our conjecture works by using the quasiparticle formula in the sudden quench \cite{Alba:2017aa, Alba:2017lvc}. 

\paragraph{(Es3)} {The quasiparticle picture is more applicable to the fast ECP rather than the slow ECP, because it is based on the sudden quench. 
However, we may slightly modify the argument by introducing the Kibble-Zurek time $t_\textrm{kz}$ \eqref{ECPkz}.  The quasiparticles are generated not at $t\sim0$ but at $t \sim t_\textrm{kz}$ in the slow ECP with large $l$, {thus  $t_c$ in the slow ECP with $z=1$ from the fast quasiparticles is}
\begin{equation} \label{tt777}
t_c \sim t_\textrm{kz} + \frac{l}{2}  \qquad (z=1)\,,
\end{equation}
{as shown in} Figure \ref{fig4b}. Indeed, it {is} confirmed by the correlator method in \cite{Nishida:2017hqd}.
By the same argument as in the slow ECP case, we {expect $t_c$ to satisfy}\footnote{{{For $z=1$ and large $l$, $t_{\textrm{kz}} +\frac{l}{2 |v_\textrm{max}|}$ is a good criteria because both $t_\textrm{kz}$ and $v_\textrm{max}$ is evaluated at $k\sim0$~\cite{Nishida:2017hqd, Fujita:2018lfj}.  However, for $z\ge2$, it may not because $t_{\textrm{kz}}$ is determined from $\omega_k(t)$ at $k=0$ while $v_\textrm{max}$ is defined at finite $k$ away from $k=0$ as shown in  Figure \ref{sk00}. In principle, we have to find $k=\tilde{k}$ such that $t_{\textrm{kz}}(\tilde{k}) +\frac{l}{2 |v(\tilde{k}|)}$ can be minimized and use it as a criteria. Here, $t_{\textrm{kz}}(k) $ is the time scale when the adiabaticity of $\omega_k(t)$ starts breaking, {of which precise meaning is shown in Eq.~(3.10) in  \cite{Fujita:2018lfj}.} Note that in principle $t_{\textrm{kz}}(k)$ can be defined for every $k$, but we used $t_{\textrm{kz}} := t_{\textrm{kz}}(0)$ for simplicity. }} \label{foo778}} 
{
\begin{equation} \label{z2011}
  t_c \gg  t_\textrm{kz} + \frac{l}{2 |v_\textrm{max}|} \,,
\end{equation}}for $z>1$.  Figure \ref{fig4b} is one of the examples.\footnote{{As we noted in \eqref{z202}, it does not guarantee $t_c >  t_\textrm{kz} + \frac{l}{2}$. However, in our cases, it turns out to be true. See section \ref{whydct} for more details.}} We {also} confirmed this delayed $t_c$ for $z=4, 6, 8$.
}

\paragraph{(Ef4,5) and (Es4,5)} The free Lifshitz scalar field theories with $z>1$ have a higher spatial derivative interaction. After discretizing these field theories to lattice theories, this higher derivative interaction becomes a long-range interaction between the fields at two separate lattice points. As explained in \cite{MohammadiMozaffar:2017nri, He:2017wla, MohammadiMozaffar:2018vmk},  $\Delta S_A$ increases as {$z$} increases because of the long-range interaction (the properties (Ef4) and (Es4)).  We suspect that the nonlinear increase of $\Delta S_A$ with $t$ described in the properties (Ef5) and (Es5) is related to the large value of $z$, but we do not have a clear interpretation.

\section{Entanglement entropy  in the CCP with $z>1$}\label{CCP}
In this section, we study the time evolution of entanglement entropy in the CCP with $z>1$ and interpret its properties.  Unlike $\Delta S_A$ in the ECP, $\Delta S_A$ in the CCP oscillates in time $t$ because of the nonzero mass potential at late times.

\subsection{Fast CCP}

As an example of the fast ($\delta t /\xi \ll 1$) CCP  with $z \ne 1$ (Lifshitz theory){,} we choose the same parameter as the fast ECP:
\begin{equation}
\xi=100\,, \quad \delta t=5\,.
\end{equation}
\begin{figure}[]
 \centering
     \subfigure[{Comparison between different subsystem sizes $l$ ($-2\le t/\xi\le40$).}]
     {{\includegraphics[width=7.2 cm]{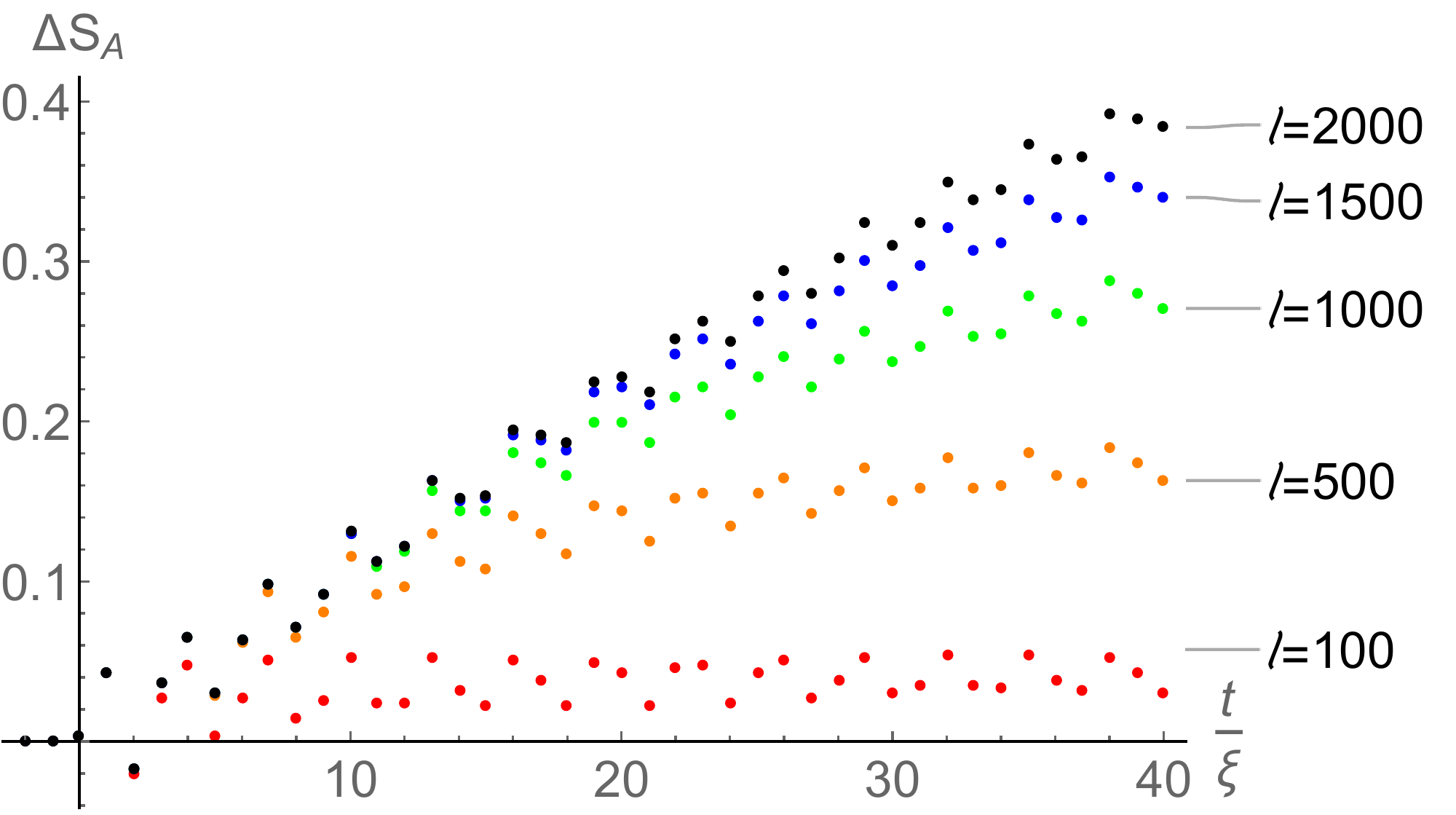} \label{fig6a}} }
          \subfigure[{Zoomed-in view} of (a) ($30\le t/\xi\le40$) to see the oscillating feature clearly.]
     {{\includegraphics[width=7.2 cm]{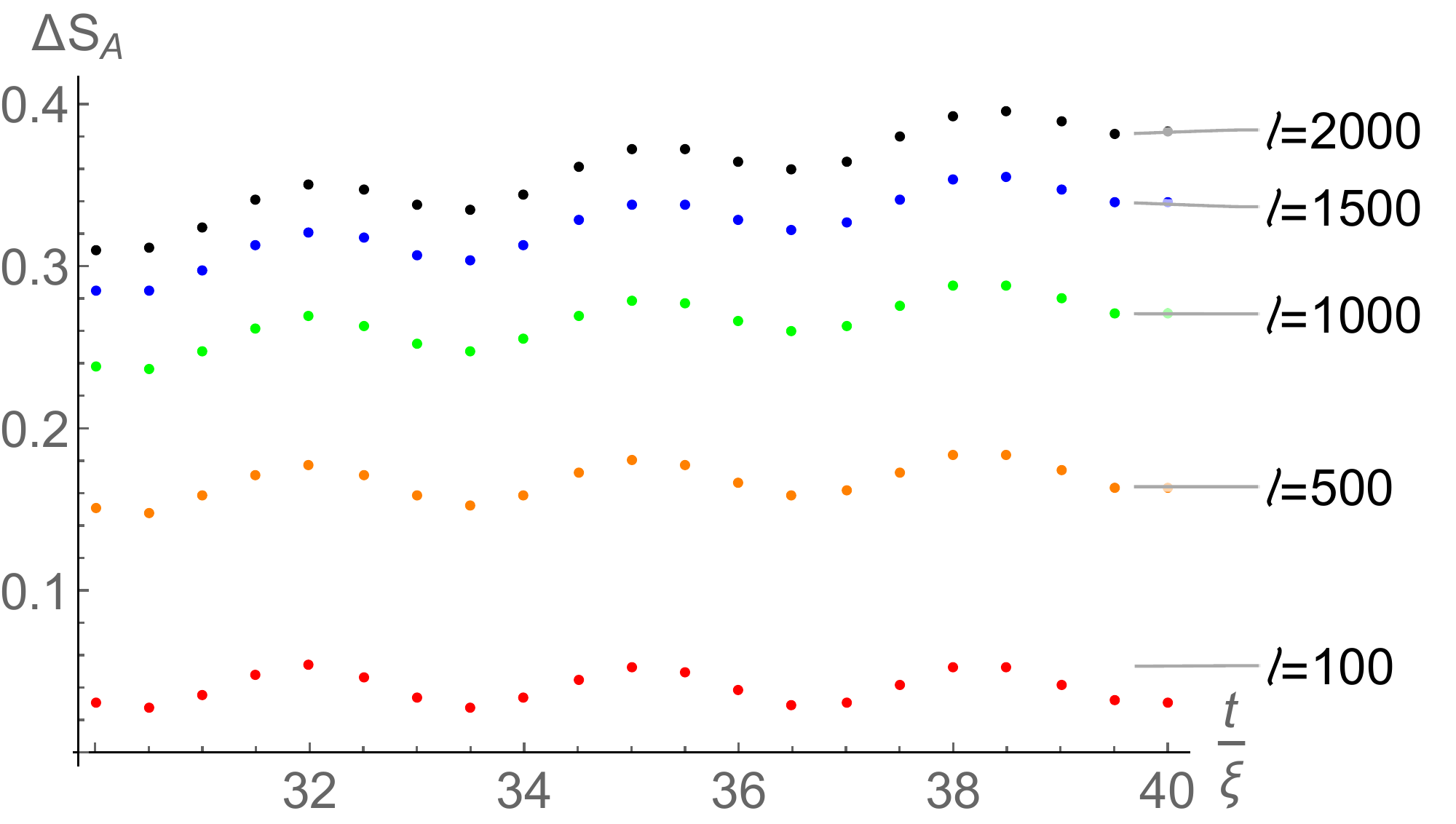} \label{fig6b}} }
     \subfigure[{Comparison between $l=1000, 2000$. The dashed line at $\frac{t}{\xi}=  \frac{l}{2\xi} =5$ for $l=1000$ is shown for comparison with $z=1$.}]
     {{\includegraphics[width=7.3 cm]{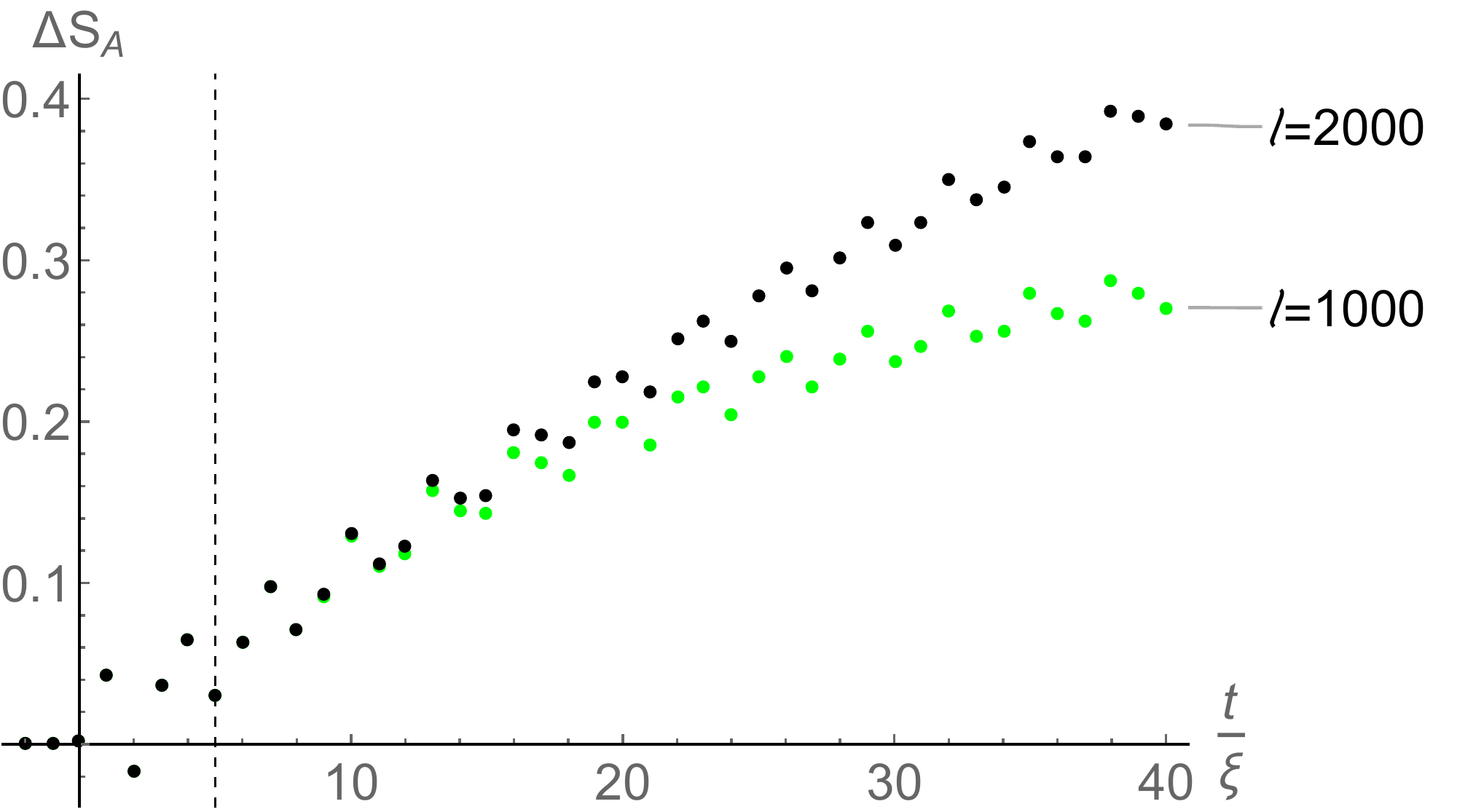} \label{fig6c}}}
           \caption{{Time dependence of $\Delta S_A$ in the fast CCP for $z=2$ ($\xi=100, \delta t=5$). } }
  \label{ccpf1}
\end{figure}
\begin{figure}[]
 \centering
     \subfigure[ {Comparison between $z=2,4$ ($-2\le t/\xi\le40$).}]
     {{\includegraphics[width=7.2 cm]{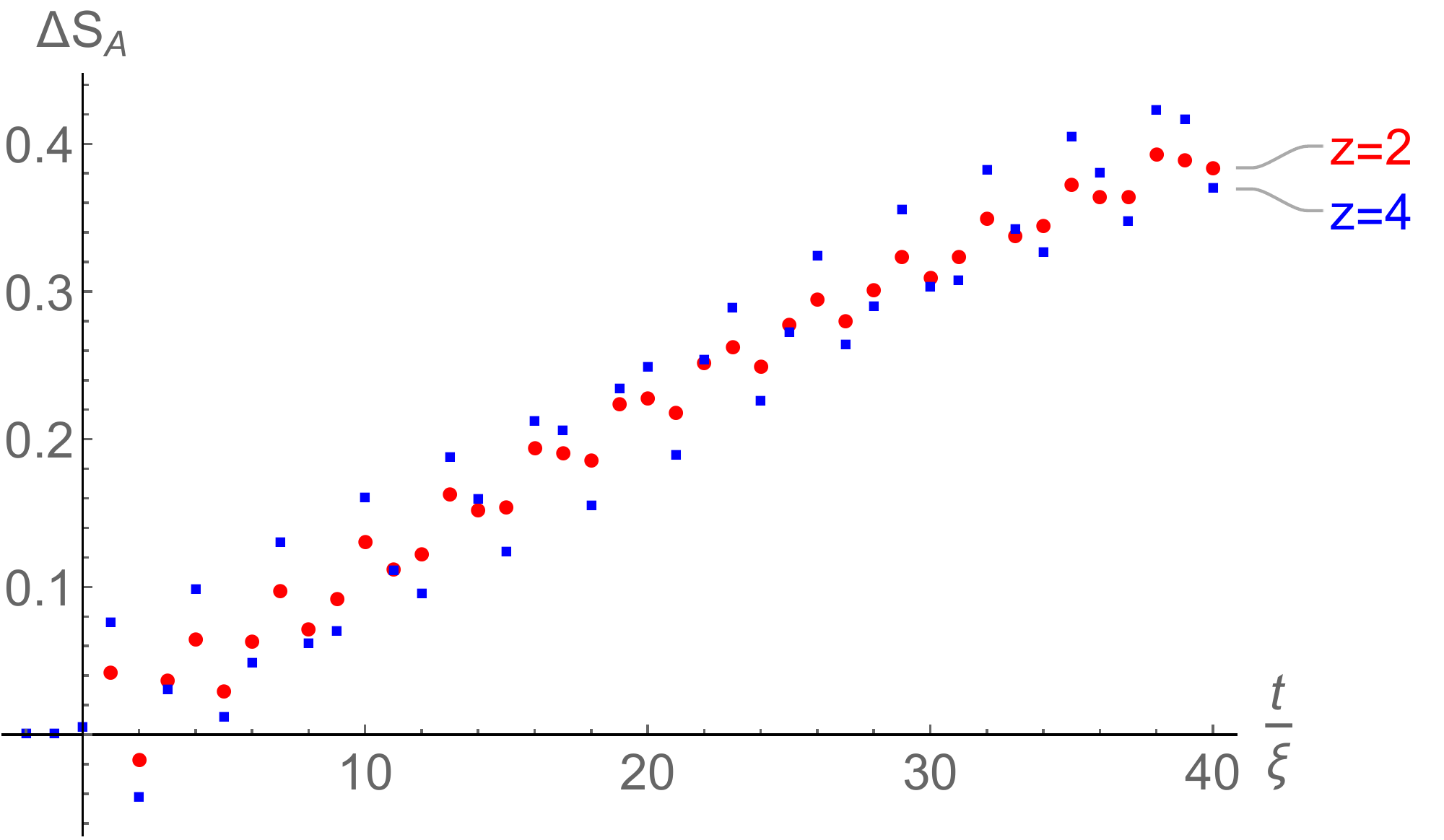} \label{fig7a}} }
          \subfigure[{Zoomed-in view} of (a) ($30\le t/\xi\le40$) to see the oscillating feature clearly.]
     {{\includegraphics[width=7.2 cm]{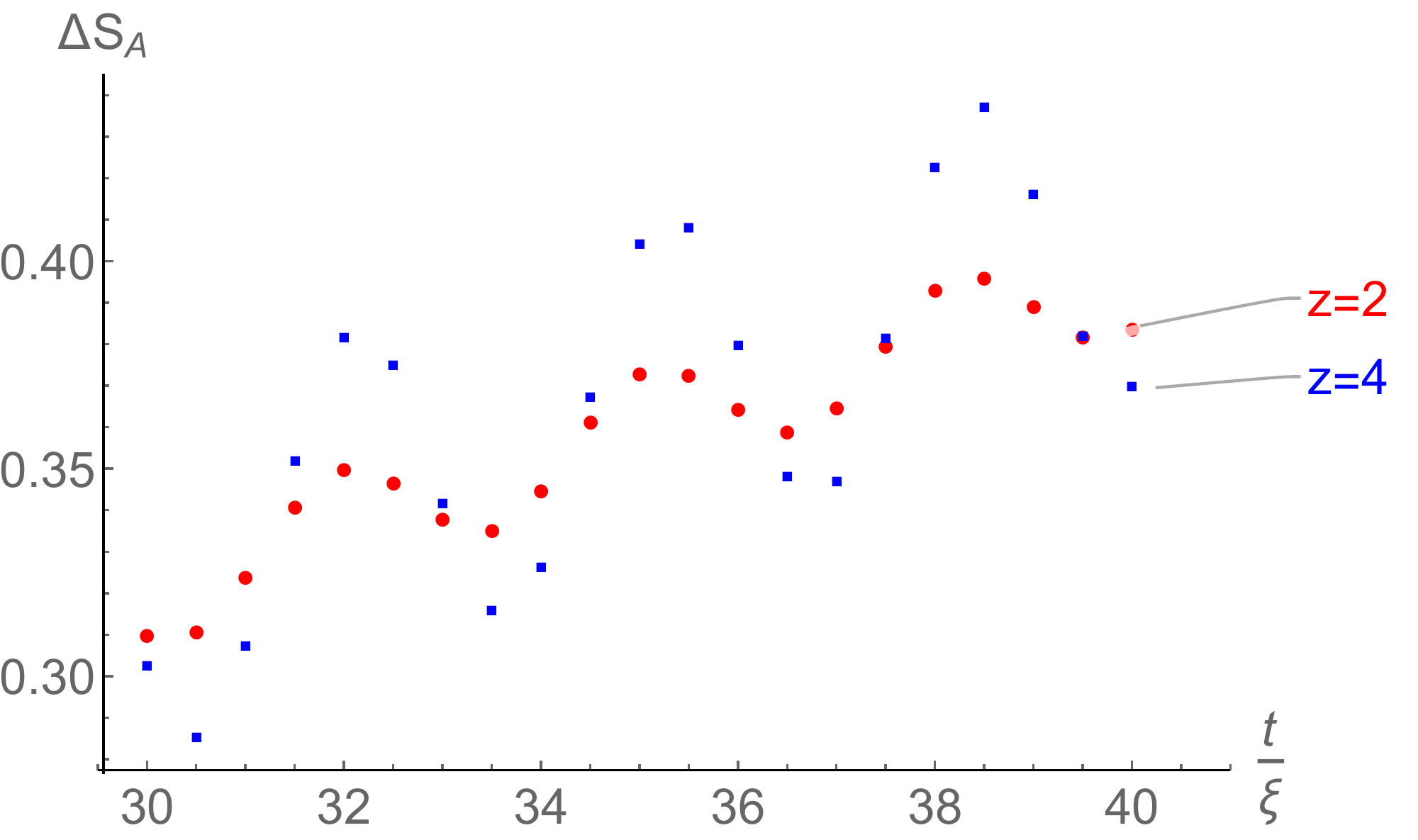} \label{fig7b}} }
              \caption{{Time dependence of $\Delta S_A$ in the fast CCP {with} $\xi=100, \delta t=5, l=2000$ and $z=2, 4$.} }
  \label{ccpf2}
\end{figure}

Figure \ref{ccpf1} shows { $l$-dependence} of $\Delta S_A$ for $z=2$ with $l=100,500,1000$, $1500,2000$. 
We observe the following properties {from} Figure \ref{ccpf1}:

\begin{enumerate}%
\item[(Cf1)] {The change} $\Delta S_A$ in the fast CCP starts  increasing at $t\sim0$ and oscillates with $t$. The period of the oscillation at late times is about $\pi \xi$. See Figure \ref{fig6b}. This period is the same as the case with $z=1$ \cite{Nishida:2017hqd}.
\item[(Cf2)] {The change} $\Delta S_A$ is {the global minimum}   around $t\sim 2\xi$ which is the same as the case with $z=1$ \cite{Nishida:2017hqd}.
\item[(Cf3)]
Like the slow and fast ECP, at early times $\Delta S_A$ has no subsystem size-dependence while at late times $\Delta S_A$ with the different subsystem sizes are different. Again, {the critical time $t_c$}  increases with the subsystem size. 
For $z=1$, it was shown that $t_c(z=1)\sim l/2$~\cite{Nishida:2017hqd}. For $z=2$, we find that $t_c(z=2) > t_c(z=1)$. For example, see Figure \ref{fig6c}, where $t_c(z=1)/\xi \sim 5$, while $t_c(z=2)/\xi > 10$ for $l=1000$.
\end{enumerate}

Figure \ref{ccpf2} shows the dynamical exponent-dependence of $\Delta S_A$ for $l=2000$ with $z=2$ and $4$.\footnote{We plot $\Delta S_A$ with $z=2, 4$ only because we need more precision for the numerical computation of $\Delta S_A$ in the fast CCP with large $z$ to reduce the  numerical error.} { The change} $\Delta S_A$ in the fast CCP shows the following properties:

\begin{enumerate}%
\item[(Cf4)] As {$z$} increases, the amplitude of oscillation in $\Delta S_A$ increases. 
\item[(Cf5)] 
The period of oscillation ($\pi \xi$) and the time scale when $\Delta S_A$ is minimum ($t\sim 2\xi$) are independent of $z$. 
\end{enumerate}

\subsection{Slow CCP}
Next, as an example of the slow ($\delta t /\xi \gg 1$) CCP  with $z \ne 1$ (Lifshitz theory){,} we choose\footnote{In the slow CCP, we use the adiabatic approximation of $f_k(t)$ and its time derivative at large $|k|$ because the computational cost of evaluating the two point functions $Q_{ab}(t)$, $P_{ab}(t)$, and $D_{ab}(t)$ increases in the slow CCP. The approximation which we use in this paper is the same one used in \cite{Nishida:2017hqd, Fujita:2018lfj}.}
\begin{equation}
\xi=10\,, \quad \delta t=1000\,.
\end{equation}

Figure \ref{ccps1} shows the subsystem size $l$-dependence of $\Delta S_A$ for $z=2$ with $l=10,50,$ $100,200,2000$. We observe the following properties in Figure \ref{ccps1}:
\begin{figure}[]
 \centering
     \subfigure[{Comparison between different subsystem sizes $l$ ($-100\le t/\xi\le200$).}]
     {{\includegraphics[width=7.2 cm]{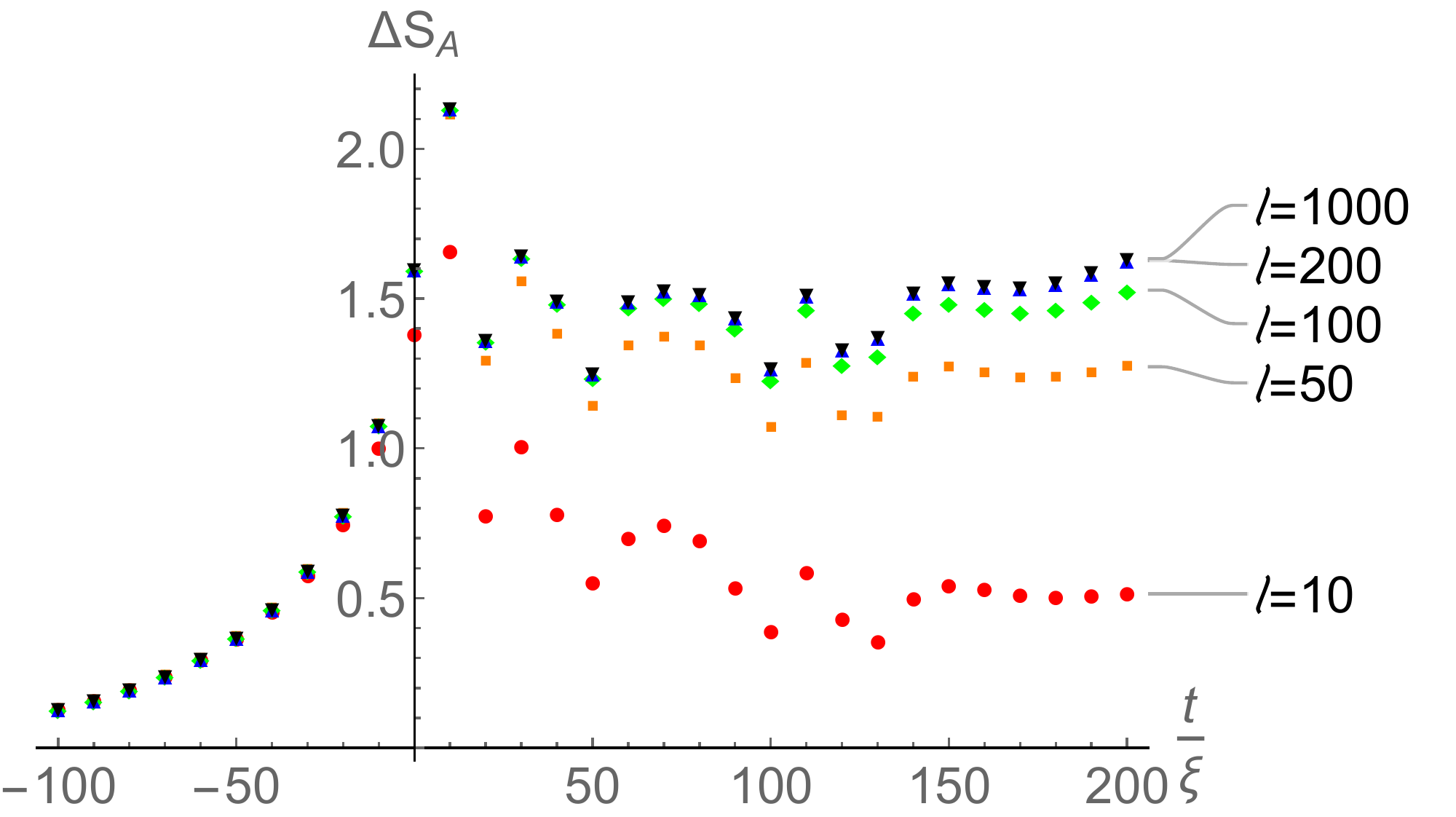} \label{}} }
          \subfigure[Zoomed-in view of (a)  ($190\le t/\xi\le200$) to see the oscillating feature clearly.]
     {{\includegraphics[width=7.2 cm]{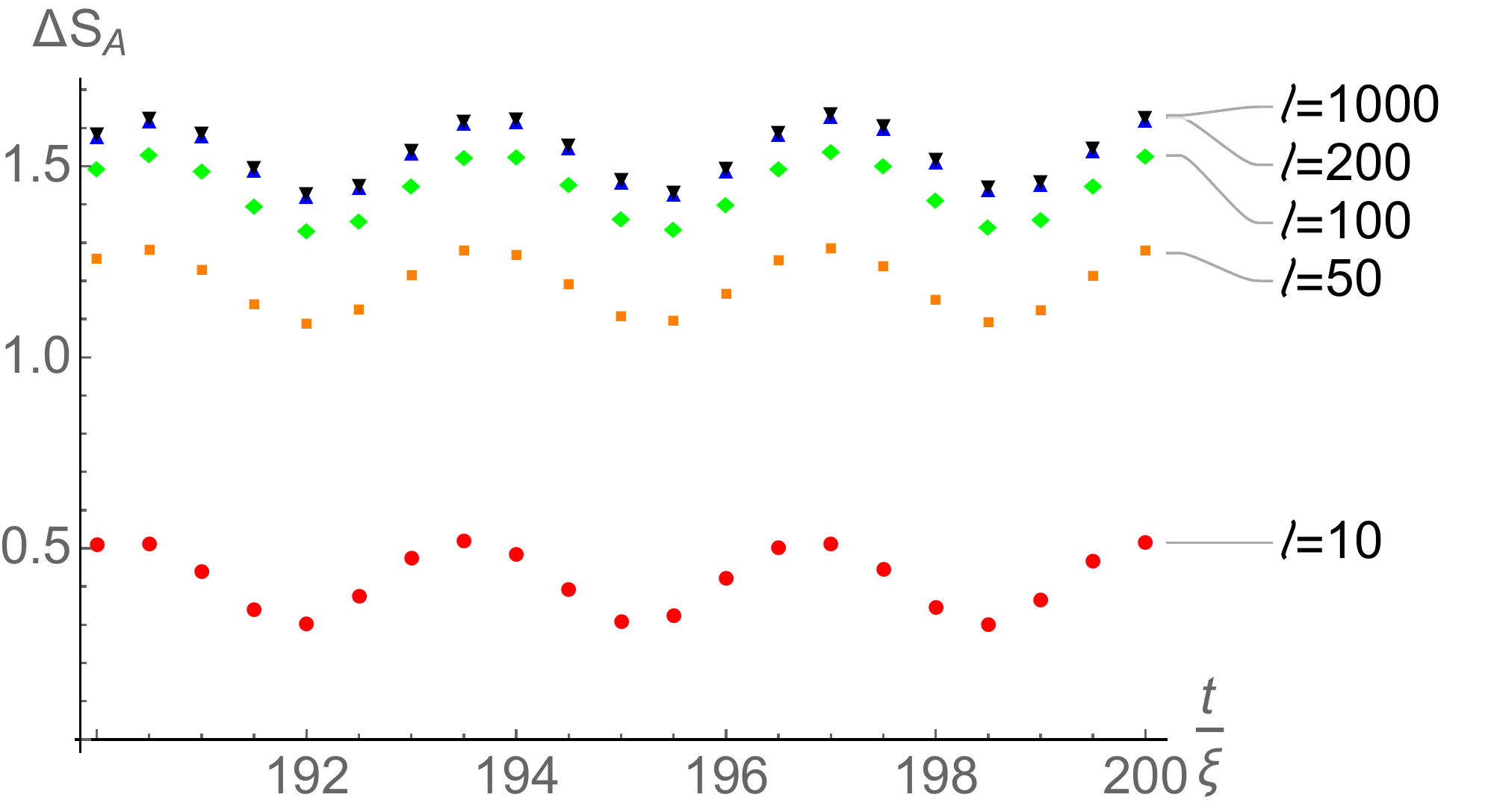} \label{}} }
        \subfigure[Zoomed-in view of (a)  ($15\le t/\xi\le25$) to see a local minimum around $t/\xi\sim 2\xi_{\textrm{kz}}/\xi\sim20$.]
     {{\includegraphics[width=7.2 cm]{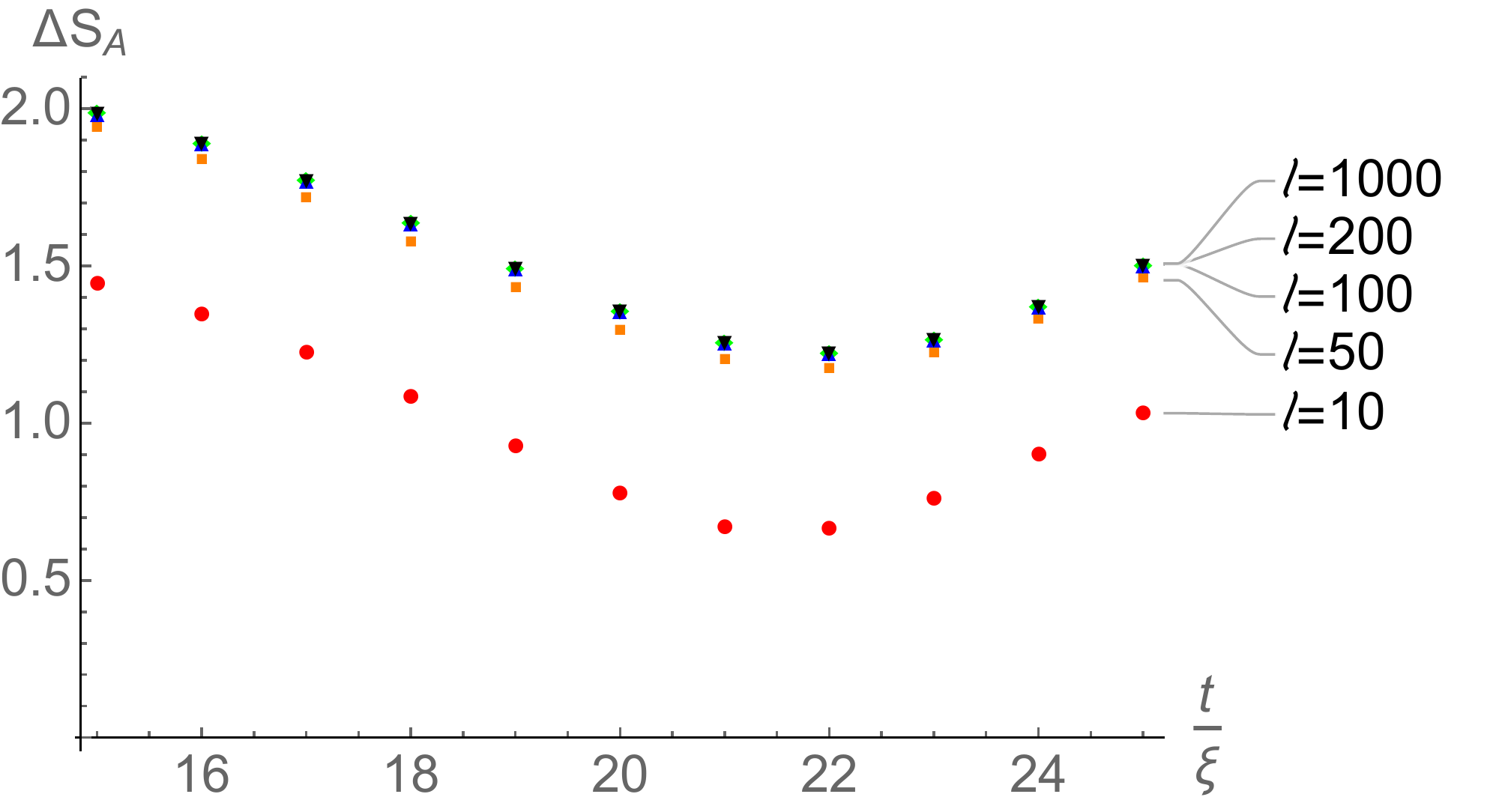} \label{}} }
              \caption{{Time dependence of $\Delta S_A$ in the slow CCP for $z=2$ ($\xi=10, \delta t=1000$).}}
  \label{ccps1}
\end{figure}

\begin{figure}[]
 \centering
     \subfigure[{Comparison between $z=2,4,6,8$  ($-100\le t/\xi\le200$).}]
     {{\includegraphics[width=7.2 cm]{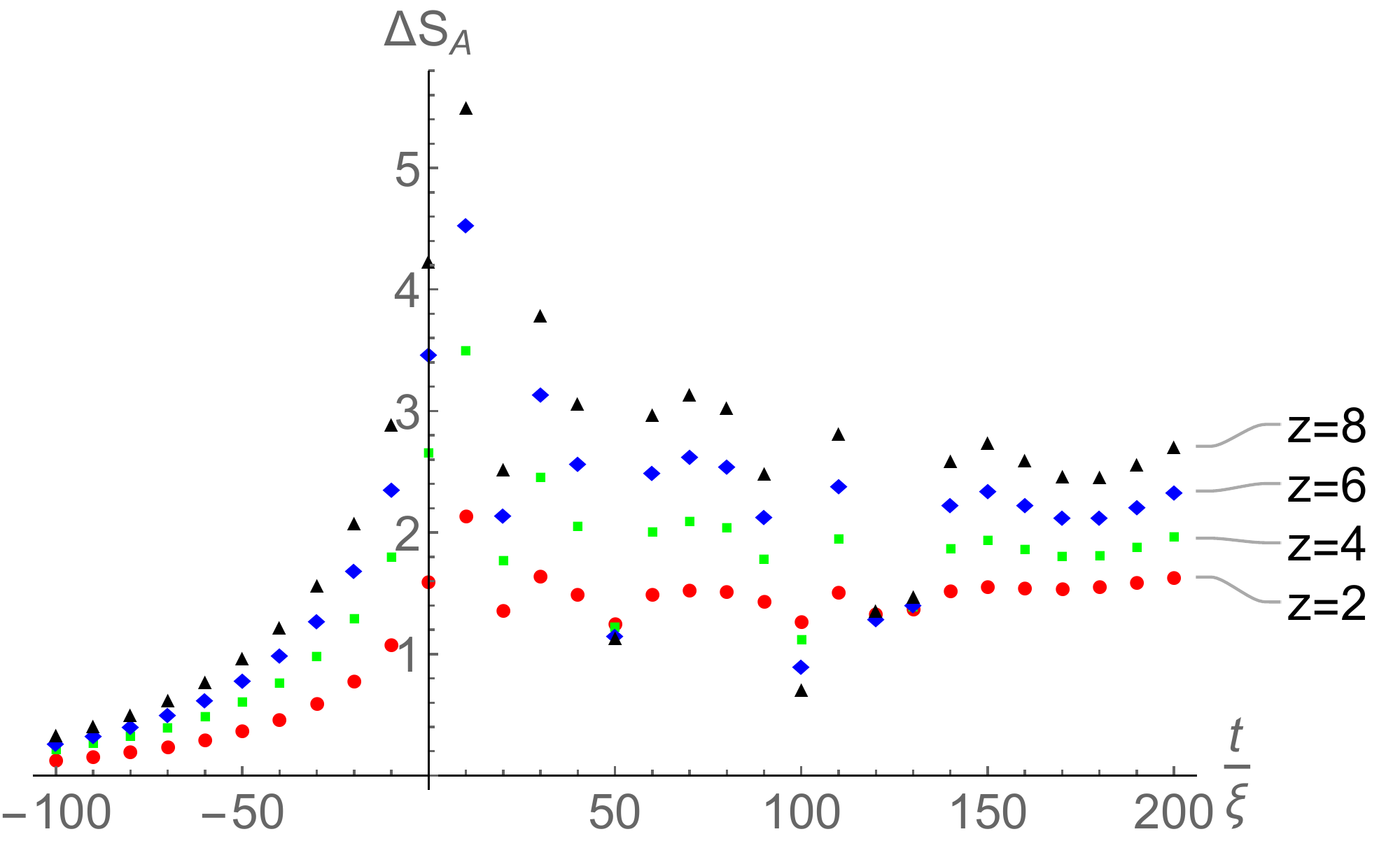} \label{}} }
          \subfigure[Zoomed-in view of (a)  ($190\le t/\xi\le200$) to see the oscillating feature clearly. ]
     {{\includegraphics[width=7.2 cm]{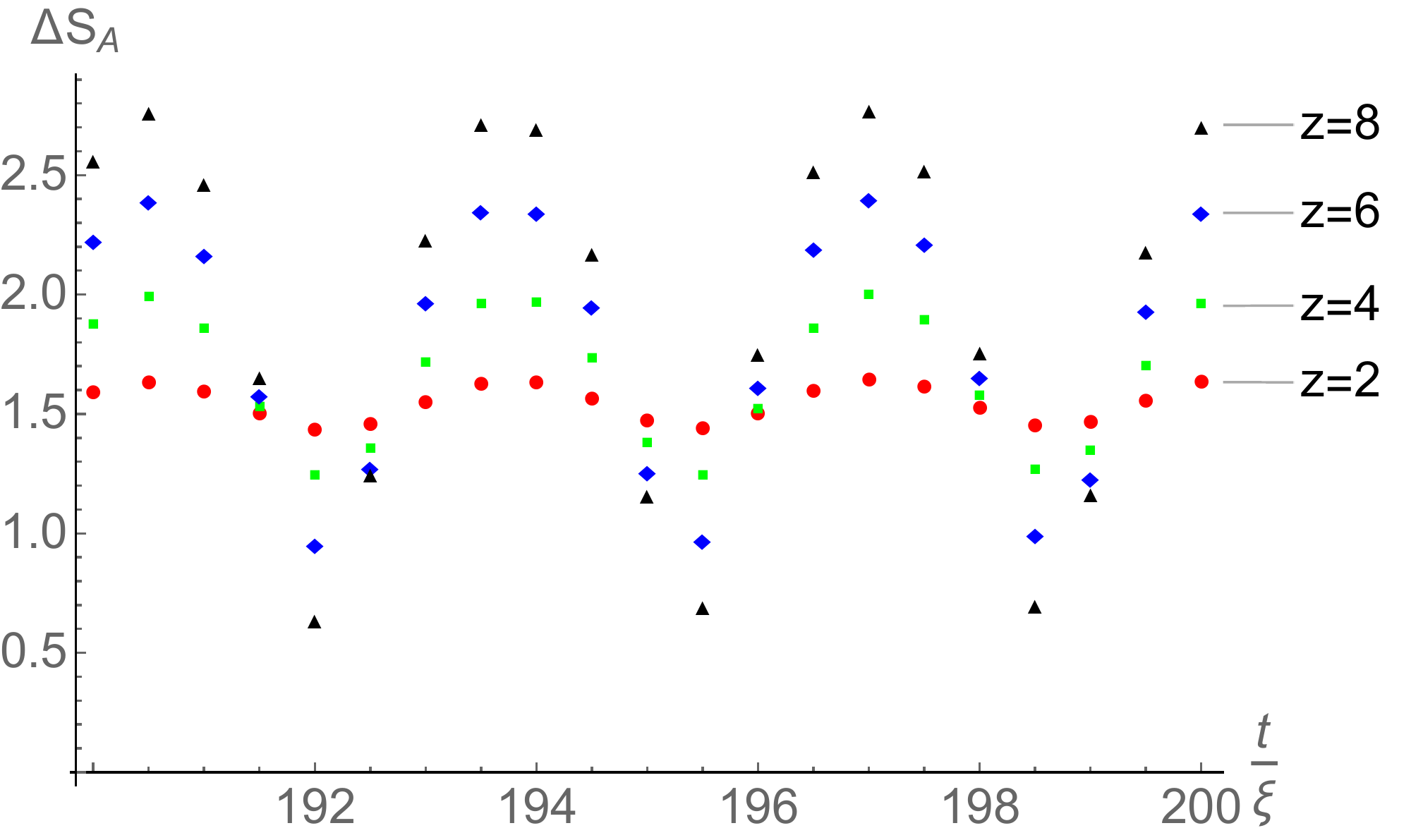} \label{}} }
     \subfigure[Zoomed-in view of (a)  ($15\le t/\xi\le25$) to see a local minimum around $t/\xi\sim 2\xi_{\textrm{kz}}/\xi\sim20$.]
     {{\includegraphics[width=7.2 cm]{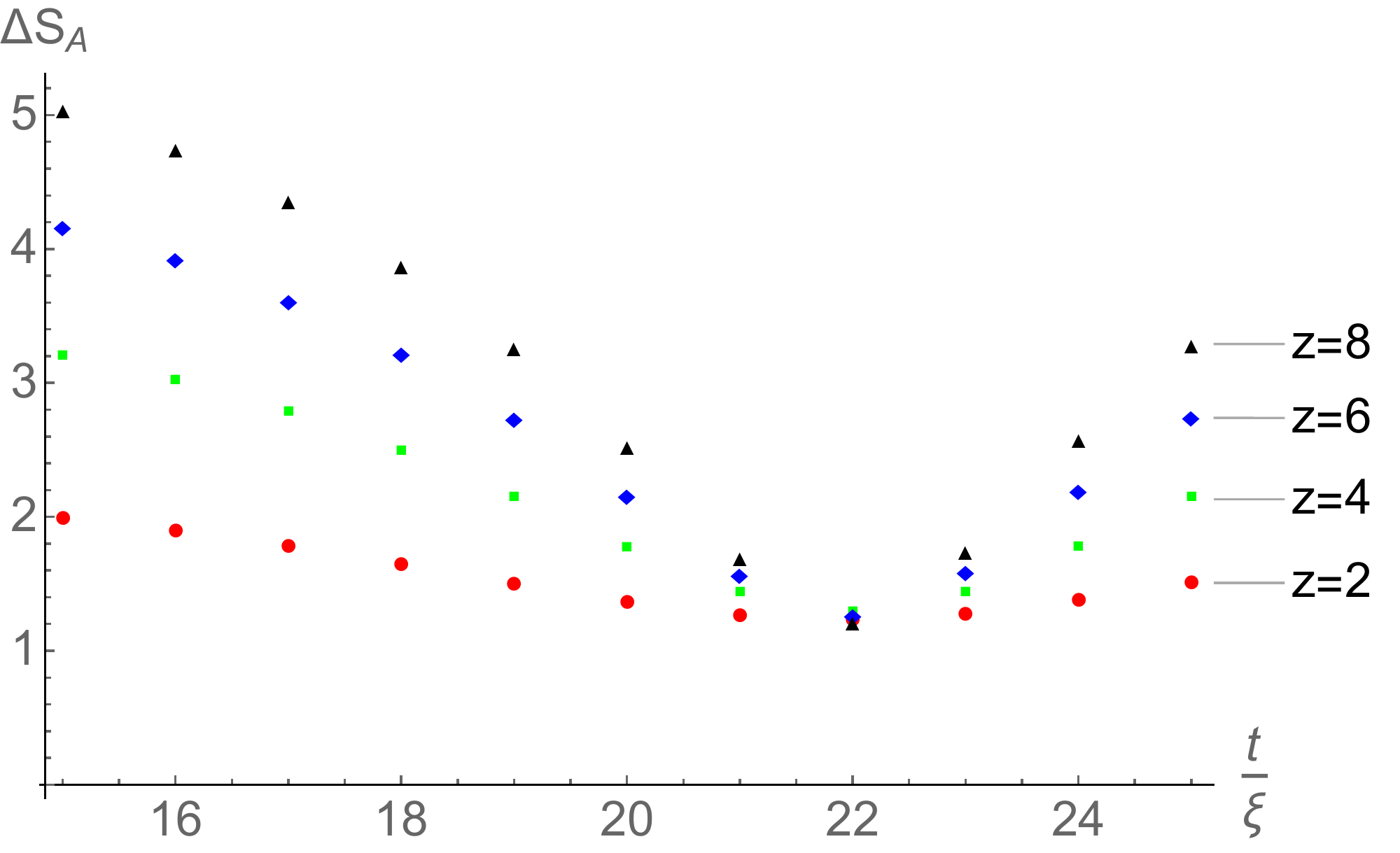} \label{}} }
              \caption{{Time dependence of $\Delta S_A$ in the slow CCP for $z=2,4,6,8$ $(\xi=10, \delta t=1000, l=1000)$.}}
  \label{ccps2}
\end{figure}
\begin{enumerate}%
\item[(Cs1)] Unlike the fast CCP case, $\Delta S_A$ in the slow CCP with $z=2$ starts increasing at  $t<0$ and oscillates with $t$. Like the fast CCP case, its period of oscillation at late times is about $\pi \xi$ which is the  same as the case with $z=1$ \cite{Nishida:2017hqd}.
\item[(Cs2)]  {The change} $\Delta S_A$ is {first} local minimum around $t\sim 2\xi_{kz}$ which is the same as the case with $z=1$ \cite{Nishida:2017hqd}.
\item[(Cs3)] Like the slow/fast ECP and fast CCP, at early times $\Delta S_A$ has no subsystem size-dependence while{,} at late times $\Delta S_A$ with the different subsystem sizes are different. Again, the critical time {$t_c$} increases with the subsystem size. 
\end{enumerate}

Figure \ref{ccps2} shows the dynamical exponent-dependence of $\Delta S_A$ for $z=2,4,6,8$ with $l=1000$. {The change} $\Delta S_A$ in the slow CCP shows the following properties:

\begin{enumerate}
\item[(Cs4)] Like the fast CCP, the amplitude of oscillation in $\Delta S_A$ becomes large when the dynamical exponent $z$ increases.
\item[(Cs5)] 
{Like the fast CCP, the period of oscillation ($\pi \xi$) and the time scale when $\Delta S_A$ is local minimum ($t\sim 2\xi_{\textrm{kz}}$) are independent of $z$.}
\end{enumerate}

\subsection{Interpretation of the properties}
In this subsection, we interpret the aforementioned properties of the entanglement entropy  for the CCP.

\paragraph{(Cf1,5) and (Cs1,5)} Like $\Delta S_A$ in the ECP, $\Delta S_A$ in the fast CCP starts increasing from $t\sim0$, while $\Delta S_A$ in the slow CCP starts increasing from $t<0$ simply because of the magnitude of $\delta t$ as explained in section \ref{IECP}.  A main difference between the ECP and CCP is that $\Delta S_A$ in the CCP oscillates in $t$ because the mass potential in the CCP at late times is nonzero. The period of oscillation in $\Delta S_A$ at late times can be inferred by \eqref{eofk} where $\omega_{k}=\sqrt{\frac{1}{\xi^2}+\left(2\sin[k/2]\right)^{2z}}$ at late times. {The period is $\frac{\pi}{\omega_k}$ and it is estimated with $k=0$ because the dominant contribution at late times comes from  small $k\sim0$. (See, for example, the entropy density plots: Figures \ref{skz1} and \ref{skz11}. {One can check that the entropy density $s(k)$ in the sudden quench is maximum at $k=0$.})} Therefore, the period is $\pi\xi$ and does not depend on $z$. 

\paragraph{(Cf3) and (Cs3)} As shown in Figure \ref{ccpf1}, the time scale at which the significant subsystem size-dependence of $\Delta S_A$ in the fast CCP with $z=2$ occurs is later than $t\sim l/2$ \footnote{In the slow CCP, $t_c$ for $z=2$  is delayed compared with $z=1$ case. However, {it} is not clear that the quasiparticle picture is valid to interpret $t_c$ even for $z=1$~\cite{Nishida:2017hqd}.} (the property (Cf3)). {In the CCP, we {also confirmed} this delayed $t_c$ for $z=4, 6, 8$.} The time scale $t\sim l/2$ is the one of $\Delta S_A$ in the fast CCP with $z=1$ and can be explained by the maximum group velocity of the quasiparticles. If we can use the quasiparticle picture to interpret the time scale of $\Delta S_A$ in the fast CCP with $z=2$, delay of the time scale with $z=2$ can be interpreted by small contribution of the fast quasiparticles to $\Delta S_A$ as explained in section \ref{IECP}.

\paragraph{(Cf4) and (Cs4)} {As $z$ increases, the long-range interaction in the Lifshitz theories} with $z>1$ seems to make the amplitude of oscillation in $\Delta S_A$ larger.

\paragraph{(Cf2,5) and (Cs2,5)} {We do not have a good understanding on why the time scales of the first local minimum of $\Delta S_A$ is around $2\xi$ and $2\xi_{\textrm{kz}}$ independently of $z$ for fast and slow CCP  respectively. This time scale is identified in \cite{Nishida:2017hqd} for $z=1$ case. }

\section{Delayed time scale: quasiparticle picture for $z>1$}\label{sqpf}
{As in} the examples in the previous sections, we found that the critical time $t_c$  for $z\ge2$ is delayed compared with the case $z=1$, {\textit{i.e.},}
\begin{equation} \label{ttt1}
 t_c(z\ge2) >  t_c(z=1)\,.
\end{equation}
In this section, we interpret this by using the the quasiparticle {formula} of entanglement entropy in the sudden quench~\cite{Alba:2017aa, Alba:2017lvc}. {We show examples of $t_c$ such that
\begin{equation} 
  t_c \gg  \frac{l}{2 |v_\textrm{max}|} \,,
\end{equation}
by using the quasiparticle formula in the sudden quench for $z\ge2$. 
This result supports our interpretation that \eqref{ttt1}, in fact, should be understood as

\begin{equation} \label{z20144}
  t_c \gg  t_\textrm{kz} + \frac{l}{2 |v_\textrm{max}|} \,,
\end{equation}
as we explained in \eqref{z201} and \eqref{z2011} in the fast and slow ECP and fast CCP for $z\ge2$.} For the fast ECP and CCP case, $t_{\textrm{kz}} \sim 0$. For the slow ECP, $t_{\textrm{kz}} $ is non-zero. See footnote \ref{foo778} for more details.
In the following section, for example, we focus on $z=2$ case.

\subsection{Review of the quasiparticle formula}


We explained a basic idea of the quasiparticle picture in section \ref{IECP}. This idea can be generalized to the case $z>1$~\cite{MohammadiMozaffar:2018vmk} and $0<z<1$ \cite{Rajabpour:2014osa, Nezhadhaghighi:2014pwa}.
We again consider a {1 dimensional} system with a subsystem $A$ of length $l$ under a {sudden} quench, where the mass potential changes at $t=0$ from the initial mass $m_0$ to the final mass $m_f$.

The idea of $z=1$ case still applies to $z>1$ case and the explanation in Figure \ref{quasip} also works for $z>1$.  {Namely in early time regime $t< \bar{t} := \frac{l}{2|v_k|}$ (yellow area) the quasiparticle pairs created in the length of $2|v_k|t$ contribute, while in late time regime $t>\bar{t}$ (green area)} the quasiparticle pairs created in the length of $l$ contribute. However, the difference between $z=1$ and $z>1$ is in the value of maximum group velocity. This is not shown in  Figure \ref{quasip}, which  describes the situation at some fixed $v_k$. 

In order to explain the delayed critical time, we need to consider the quasiparticle picture in more detail, quantitatively.
The group velocity $v_k$ is a function of $k$ and the created quasiparticle entropy density {$s(k)$, which we will explain later,} is also a function of $k$. Thus, in total, the entanglement entropy created by the quasiparticle pairs ($\Delta S^\textrm{q}_A(t)$)  {reads}~\cite{Alba:2017aa, Alba:2017lvc}:
\begin{align}
\Delta S^\textrm{q}_A(t)=t\int_{2|v_k|t<l} dk s(k)2|v_k| +l \int_{2|v_k|t>l}dk s(k) \,,\label{qpf}
\end{align}
where {$k \in [-\pi, \pi]$, and the superscript $\textrm{q}$ stands for the `quasiparticle formula' to emphasize the difference with $\Delta S_A(t)$ by the `correlator method'}. The first term comes from the yellow 
{area}, and the second term comes from the green {area} in Figure \ref{quasip}.
{Note that $\Delta S^\textrm{q}_A$ starts depending on the subsystem size $l$ after $t=\frac{l}{2|v_\textrm{max}|}$ because the second term in (\ref{qpf})  is zero before $t=\frac{l}{2|v_\textrm{max}|}$, where $v_\textrm{max}$ is the maximum group velocity of $v_k$.} 
{At fixed $t$, we can choose large but finite $l$ such that the second term in (\ref{qpf}) becomes zero. Based on this property, we define $\Delta S^\textrm{q}_A(t)|_{l\to\infty}$ as
\begin{align}
\Delta S^\textrm{q}_A(t)|_{l\to\infty}:=t\int_{-\pi}^{\pi} dk s(k)2|v_k|.\label{QPFli}
\end{align}
}

We also assume that the entropy density $s(k)$ for the entanglement entropy (\ref{qpf}) is equivalent to the thermodynamic entropy density which is computed from a density matrix $\rho_{\textrm{GGE}}$ of a generalized Gibbs ensemble \cite{Alba:2017lvc, Calabrese:2007rg} as
\begin{align}
\rho_{\textrm{GGE}}= Z^{-1}e^{-l\int \frac{dk}{2\pi}\lambda_k\hat{n}_k},
\end{align} 
where $Z$ is a normalization factor, $\lambda_k$ are Lagrange multiplies, and $\hat{n}_k=a^\dagger_k a_k$ are number operators for Hamiltonian after the quench. This assumption implies the entanglement entropy becomes the thermodynamic entropy at late time limit. Requiring  the conservation of the expectation value of the number operator {between the initial state and the generalized Gibbs ensemble at late times, $\textrm{Tr}\left[\hat{n}_k\rho_{\textrm{GGE}}\right]=\langle0|\hat{n}_k|0\rangle$}, where $|0\rangle$ is the initial ground state of Hamiltonian before the quench, we obtain 
\begin{align}
e^{\lambda_k}=1+\frac{1}{\langle0|\hat{n}_k|0\rangle}.
\end{align}

In free scalar theories in the sudden quench, the explicit form of $s(k)$ is \cite{Alba:2017lvc, Calabrese:2007rg}
\begin{align}
s(k)=&\frac{1}{2\pi}\left[(n_k+1)\log(n_k+1)-n_k\log(n_k)\right],\label{ed}\\
n_k:=&\langle0|\hat{n}_k|0\rangle=\frac{1}{4}\left(\frac{\omega_k}{\omega_{0, k}}+\frac{\omega_{0, k}}{\omega_{k}}\right)-\frac{1}{2},\label{nk}
\end{align}
where we use the dispersion relations $\omega_{0, k}$ before the quench and $\omega_{k}$ after the quench of the Lifshitz theories as
\begin{align}
\omega_{0, k}=\sqrt{m_0^2+\left(2\sin[k/2]\right)^{2z}}, \;\;\;\omega_{k}=\sqrt{m_f^2+\left(2\sin[k/2]\right)^{2z}}.\label{dr}
\end{align}
The group velocity $v_k$ after the quench in these theories is
\begin{align}
v_k:=\frac{d \omega_{k}}{d k}=\frac{z \cos[k/2] \left(2\sin[k/2]\right)^{2z-1}}{\sqrt{m_f^2+\left(2\sin[k/2]\right)^{2z}}}.\label{gv}
\end{align}
With these expressions (\ref{ed}), (\ref{nk}), (\ref{dr}), and (\ref{gv}), one can compute (\ref{qpf}) explicitly.

\subsection{Examples}
 
Entanglement entropy with $z>1$ by the quasiparticle picture was also studied in \cite{MohammadiMozaffar:2018vmk}, where $m_0=1$ and {$m_f=0,2^z$} are considered. Compared with \cite{MohammadiMozaffar:2018vmk}, we are interested in small $m_0 \ll 1$ because i) it corresponds to the field theory limit, \textit{i.e.}, $\xi \gg 1$ where $\xi$ is measured by the lattice {spacing}; ii) it corresponds to our model in section \ref{FECP1}. Furthermore, our analysis is extended to explain the delayed $t_c$, which is complementary to \cite{MohammadiMozaffar:2018vmk}.

For example, let us consier $\Delta S^\textrm{q}_A$ for $z=2$.  We choose a small but nonzero value of $m_f=10^{-6}$ to avoid the divergence of $n_k$ at $k=0$. Figure \ref{QPF1} plots $\Delta S^\textrm{q}_A(t)$ in \eqref{qpf} for various subsystem sizes $l=100,200,300,400,500,1000,2000$. Figure \ref{fig11a}  is for the initial mass $m_0=0.01$, and Figure \ref{fig11b}  is for $m_0=1$. 
\begin{figure}[]
 \centering
     \subfigure[$m_0=10^{-2}$]
     {{\includegraphics[width=7.2 cm]{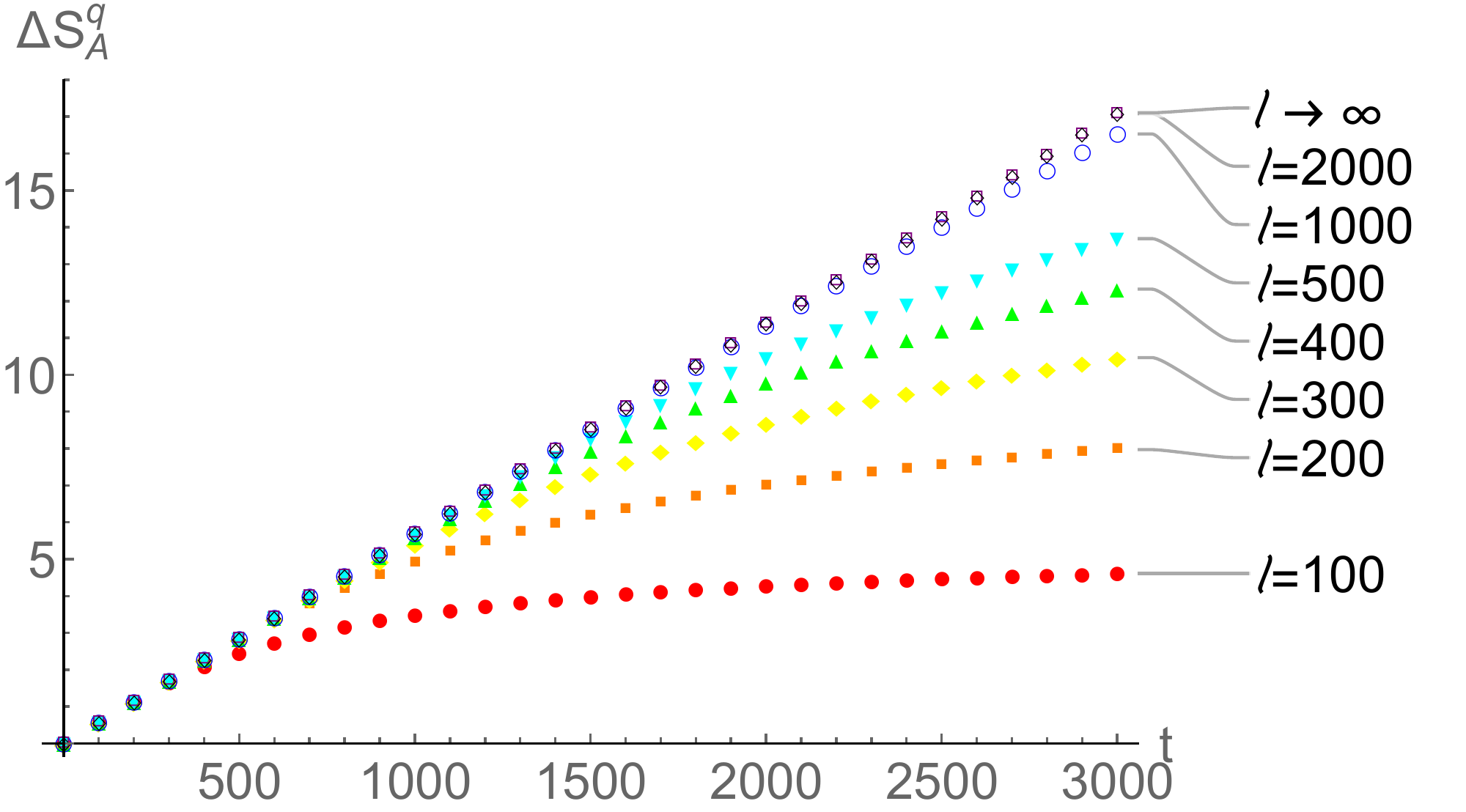} \label{fig11a}} }
     \subfigure[$m_0=1$]
     {{\includegraphics[width=7.2 cm]{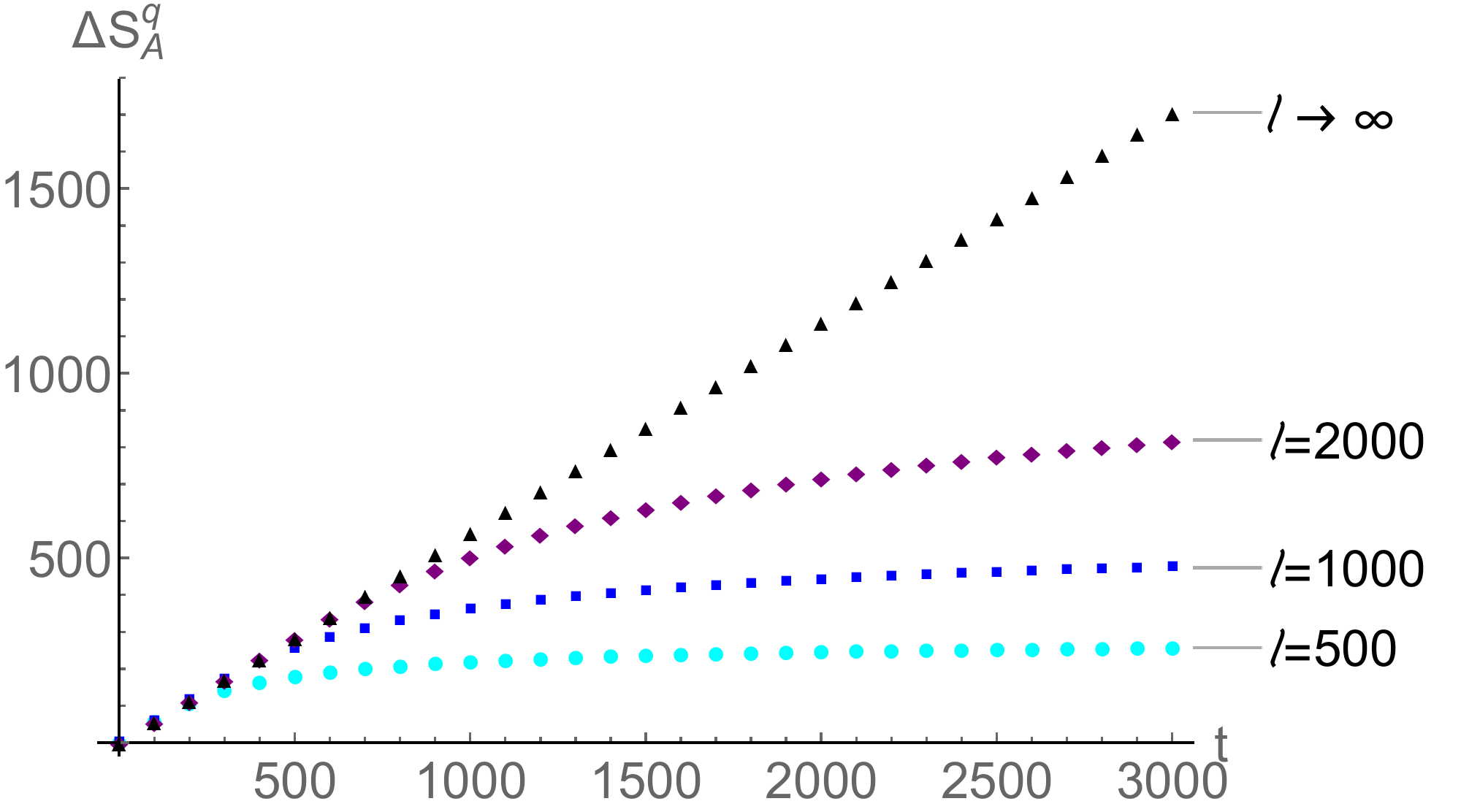} \label{fig11b}}}
           \caption{{The change of the entanglement entropy $\Delta S^\textrm{q}_A$ (\ref{qpf}) in the sudden quench for $z=2$ and $m_f=10^{-6}$.  The change $\Delta S^\textrm{q}_A$ for $l\to\infty$ is (\ref{QPFli}).} }
  \label{QPF1}
\end{figure}

Note that Figure \ref{fig11a} should be compared with Figure \ref{fig2a} {because the sudden quench is a limit of the fast ECP. For both plots, the initial mass is the same, and the finial mass is almost zero. However, Figure \ref{fig11a} is computed by the quasiparticle picture \eqref{qpf} while Figure \ref{fig2a} is computed by the correlator method.} (The horizontal time axis is scaled by $\xi$ in Figure \ref{fig2a}.) {Their significant subsystem size-dependences agree with each other very well}\footnote{{The numerical values of the entanglement entropy is slightly different. Roughly speaking, 
\begin{equation}
\Delta S_A  ({\text{Figure 2(a)}}) \ge \Delta S^\textrm{q}_A ({\text{Figure 11(a)}}) \,.
\end{equation}
This is because the mass ratio $m_0/m_f$ is too big as argued in \cite{MohammadiMozaffar:2018vmk}.
(See  Figures 6 and 8  in \cite{MohammadiMozaffar:2018vmk}.)} \label{f12}}.

\subsection{Why delayed critical time?} \label{whydct}

Let us now turn to our main question: why is the critical time $t_c$ delayed for $z=2$ compared with the $z=1$ case? To answer the question, we first revisit the argument for $z=1$ {(See \eqref{tt7}).
In order to determine $t_c$, we use the maximum group velocity $v_{\textrm{max}}$ in \eqref{tt7}. For the massless quasiparticle with $z=1$,  we obtain $v_{\textrm{max}} =1$ and $t_c\sim\frac{l}{2|v_{\textrm{max}}|}=\frac{l}{2}$. }

To investigate this property in more detail, let us rewrite (\ref{qpf}) as
\begin{align}
\Delta S^\textrm{q}_A(t)=t\int_{-\pi}^{\pi} dk s(k)2|v_k| -\int_{2|v_k|t>l}dk (2|v_k|t-l)s(k)\,.\label{qpf2}
\end{align}
Here, the first term in (\ref{qpf2}) does not depend on $l$. The change $\Delta S^\textrm{q}_A$ depends on $l$ after $t=\frac{l}{2|v_\textrm{max}|}$ because of the second term in (\ref{qpf2}). Only after $t \sim \frac{l}{2|v_\textrm{max}|}$,  the quasiparticle pairs with $v_k \sim v_{\textrm{max}}$ starts contributing.  However, in this case, the factor $(2|v_k|t-l)$ in the integrand of the second term is small. Unless $s(k)$ is {large} enough the $l$-dependence due to the second term will be negligible near $t \sim \frac{l}{2|v_\textrm{max}|}$ even though it is non-zero. Thus, we find that a naive argument for $z=1$ needs to be revisited.

\paragraph{$z=1$ case}
In Figure \ref{skz1}, we make plots of (\ref{ed}) and (\ref{gv}) for $s(k)$ and $v_k$ respectively, where  $m_0=10^{-2}$ and  $m_f=10^{-6}$. 
The group velocity $v_k$ is maximum ($v_{\textrm{max}}\sim1$) near $k\sim0$ (Figure \ref{fig12a}).  Near $k\sim0$, $s(k)$ is dominant, which makes the integrand of the second term of \eqref{qpf2} big enough as we suspected. It explains $t_c\sim\frac{l}{2}$ for $z=1$. 
\begin{figure}[]
 \centering
     \subfigure[Group velocity $v_k$. The right panel is a zoomed-in view of the left panel near $k=0$.   ] 
     {{\includegraphics[width=6 cm]{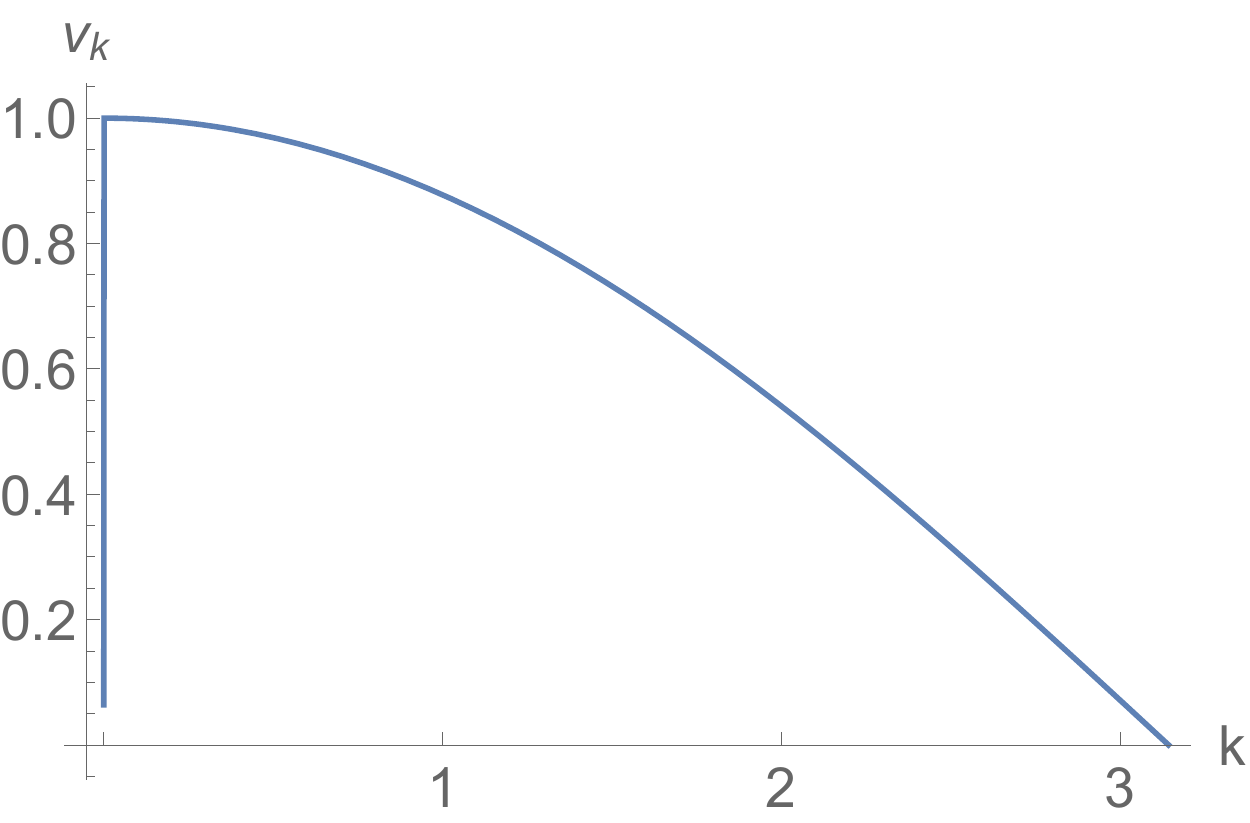} } 
      {\includegraphics[width=6 cm]{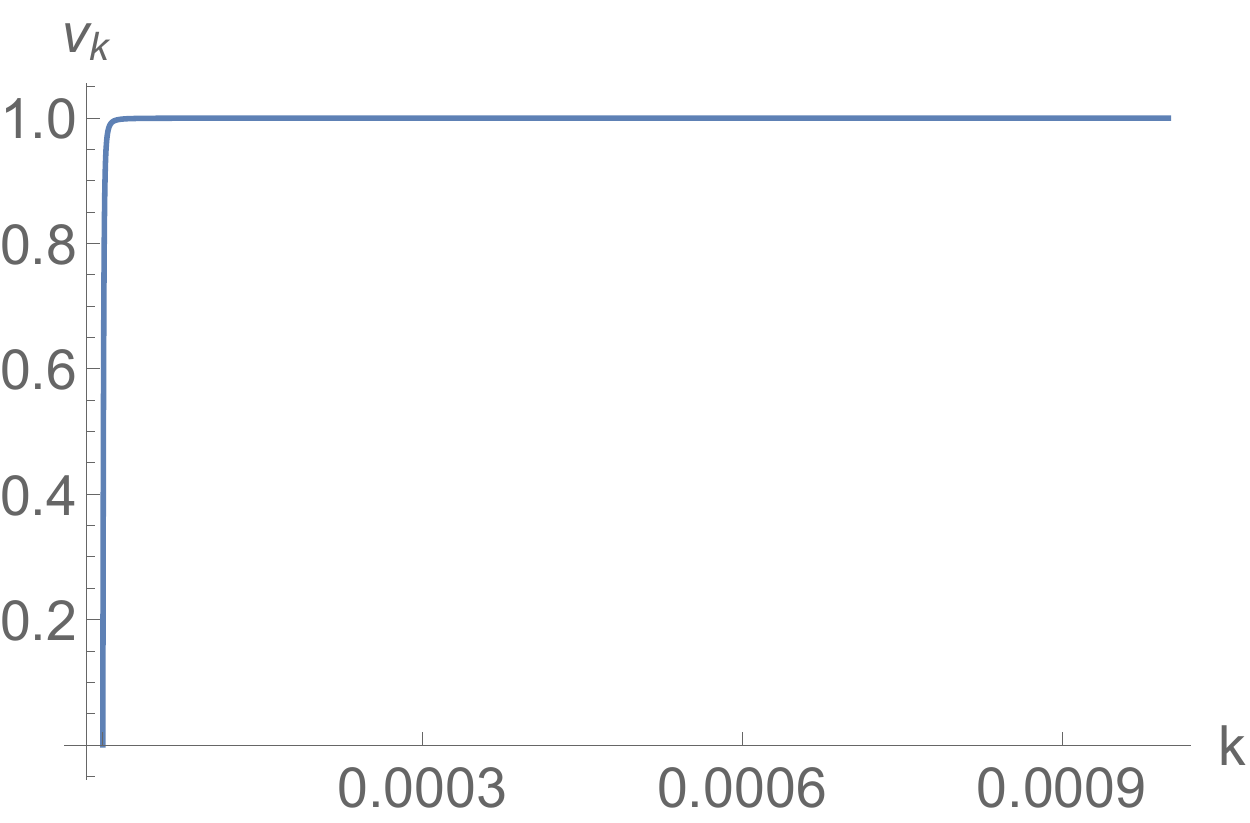} } \label{fig12a}}
     \subfigure[Entropy density $s(k)$. The right panel is a zoomed-in view of the left panel near $k=0$.] 
     {{\includegraphics[width=6 cm]{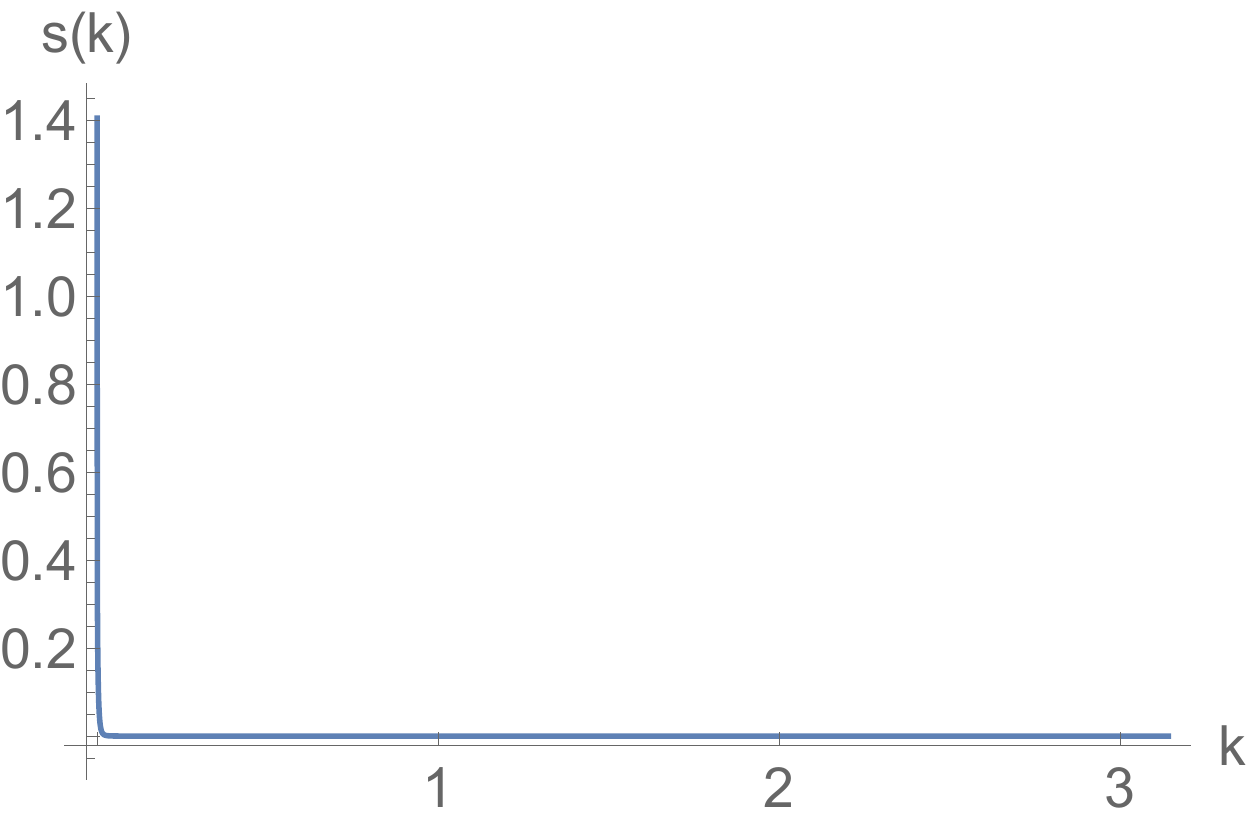} }
          {\includegraphics[width=6 cm]{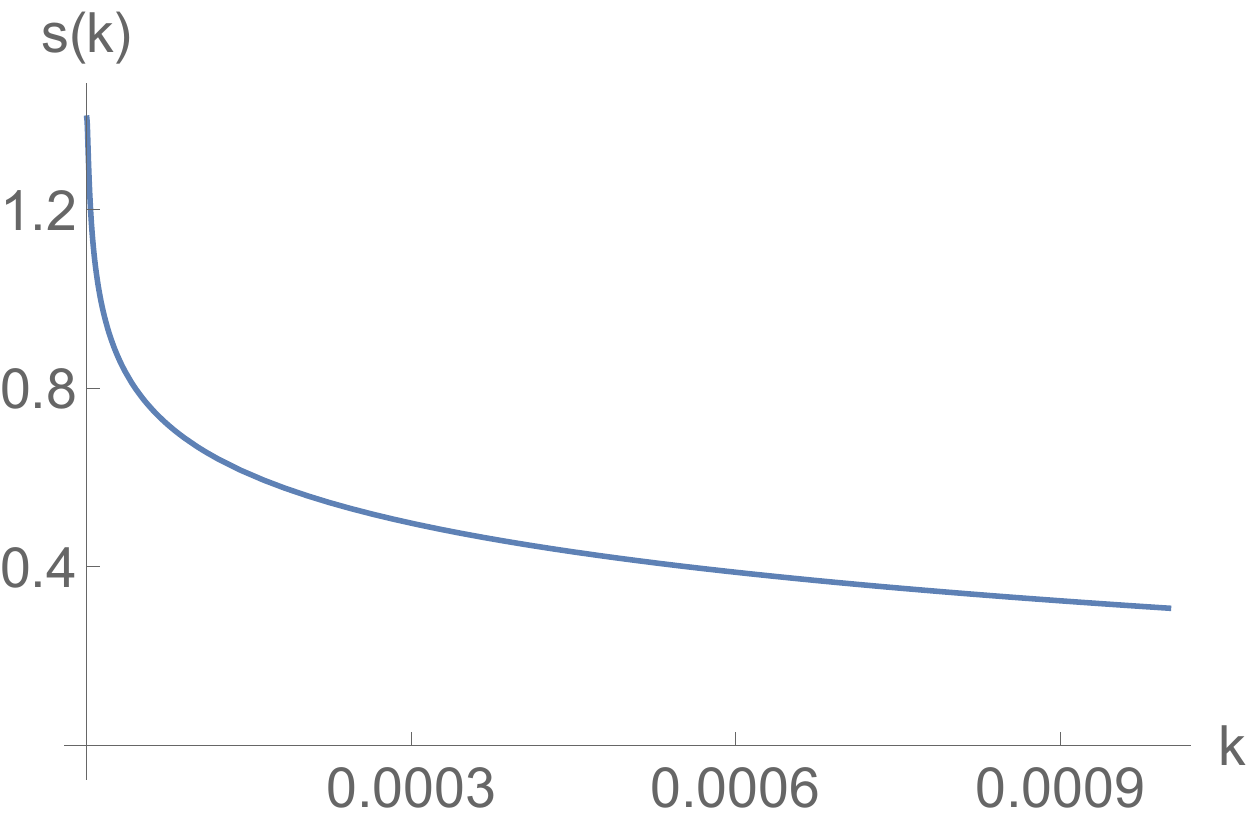}} \label{fig12b}}
           \caption{{Group velocity $v_k$ and entropy density $s(k)$ for $z=1$, $m_0=10^{-2}$, and  $m_f=10^{-6}$.}  }
  \label{skz1}
\end{figure}

\paragraph{$z=2$ case}
Let us turn to the $z=2$ case. 
Figure \ref{sk00} shows  the group velocity $v_k$ for $m_0=10^{-2}$. Unlike the $z=1$ case, $|v_k|$ at $|k|\sim 1.6$ is maximum, which is away from $k=0$. 
Figure \ref{fig14a} shows the quasiparticle pair entropy density $s(k)$ of the fast quasiparticles around $|k|\sim 1.6$ is much smaller than the one of the slow quasiparticles around $k\sim0$. {The smallness of $s(k)$ of the fast quasiparticles makes their contribution to the entanglement entropy small} (the integrand of the second term of \eqref{qpf2} is small).  {Consequently, $t_c$ becomes delayed compared with $t=\frac{l}{2|v_\textrm{max}|} \sim \frac{l}{4}$. Thus, for example, $t_c > 3000$ for $l=1000$ from Figure \ref{fig11a}, \textit{i.e.},
\begin{equation}\label{ooo1}
t_c(m_0 = 10^{-2}, l=2000) > 3000 > \frac{l}{4} = 250 \,.
\end{equation}

{For another comparison, we show the entropy density $s(k)$ for $m_0=1$ in Figure \ref{fig14b}. 
The entropy density $s(k)$ for the fast quasiparticles with $m_0=1$ is larger than that with $m_0=10^{-2}$. The group velocity of quasiparticles in the sudden quench with $m_0=1$ is  the same as the one with $m_0=10^{-2}$ as in Figure \ref{sk00}.  Thus, we expect $t_c$ to be less delayed compared with the $m=10^{-2}$ case. Indeed, it turns out to be the case.} 
For example,  $t_c \sim 800$ for $l=2000$ from Figure \ref{fig11b}, \textit{i.e.},
\begin{equation}\label{ooo2}
t_c(m_0 = 1, l=2000) \sim 800 > \frac{l}{4} = 250 \,,
\end{equation}
which corresponds to the case \eqref{z202} because
\begin{equation}
t_c(m_0 = 1, l=2000) \sim 800 < \frac{l}{2} = 1000 \,.
\end{equation}
In short, in this case, $t_c$ is not delayed so much compared with $z=1$ case, but still delayed compared with $\frac{l}{2|v_\textrm{max}|}$.}

\begin{figure}[]
 \centering
     {\includegraphics[width=6 cm]{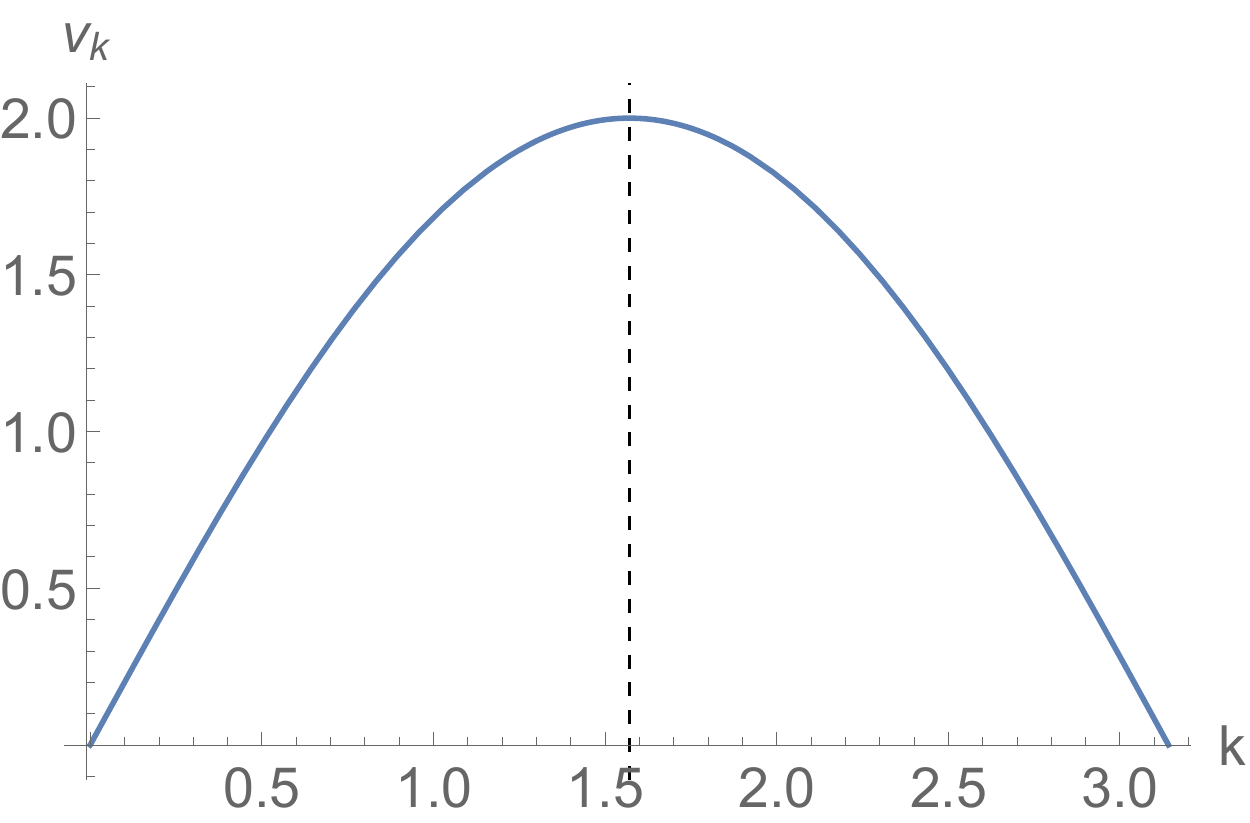} \label{FIGnormal1a}}
           \caption{Group velocity $v_k$ for $z=2$ and the final mass  $m_f=10^{-6}$. It is independent of the initial mass $m_0$. The group velocity $v_k$ at $k = k^* \approx 1.6$ is maximum. }
  \label{sk00}
\end{figure}
\begin{figure}[]
 \centering
     \subfigure[$m_0=10^{-2}$. The right panel is a zoomed-in view of the left panel to show the value of $s(k^*)$.]
     {{\includegraphics[width=6 cm]{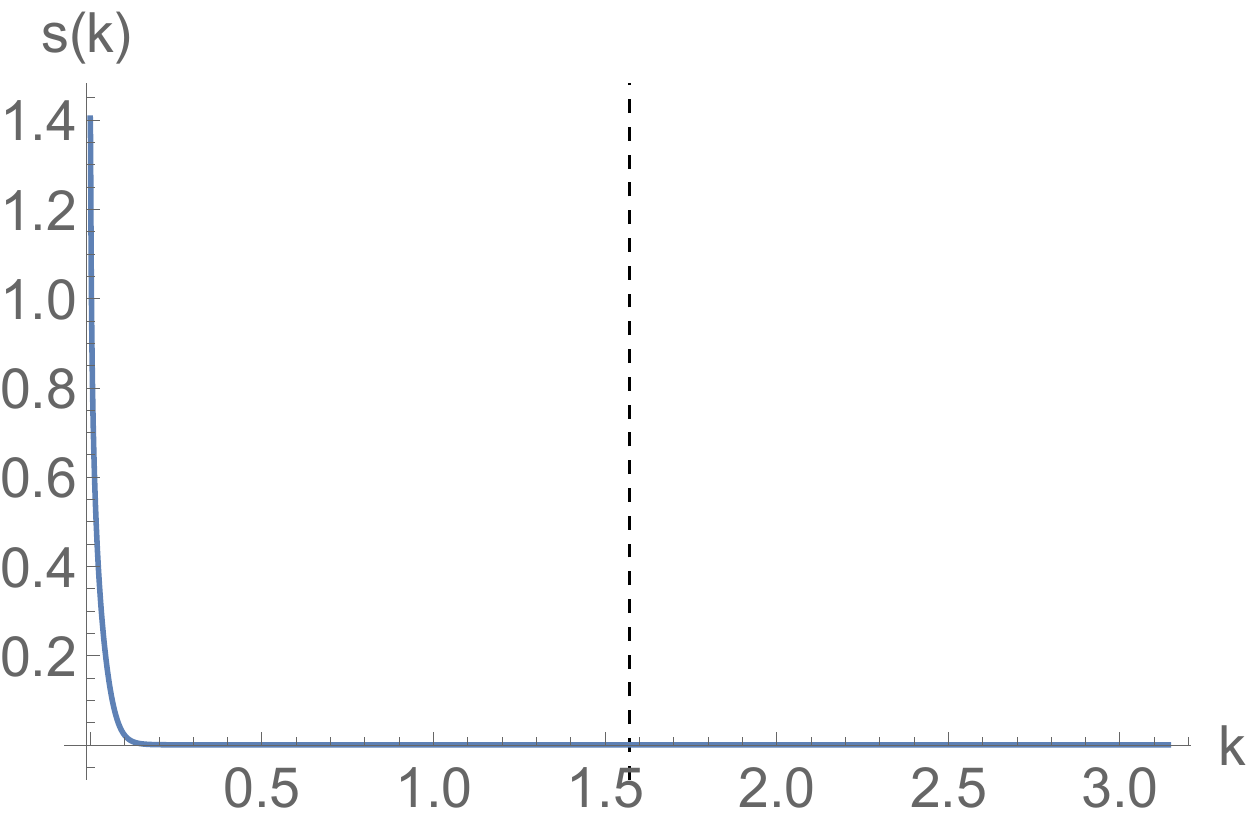} \label{fig14a}}
          {\includegraphics[width=6 cm]{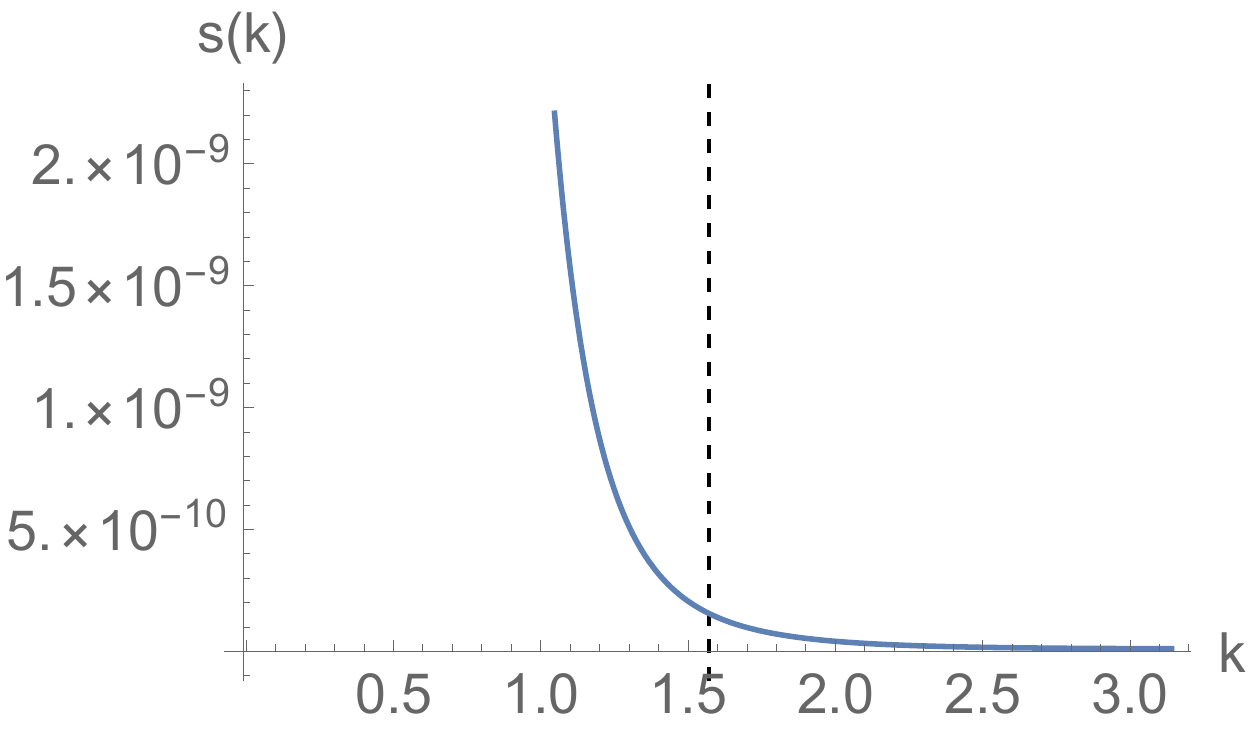} \label{}}}
               \subfigure[$m_0=1$. The right panel is a zoomed-in view of the left panel to show the value of $s(k^*)$.   ]
     {{\includegraphics[width=6 cm]{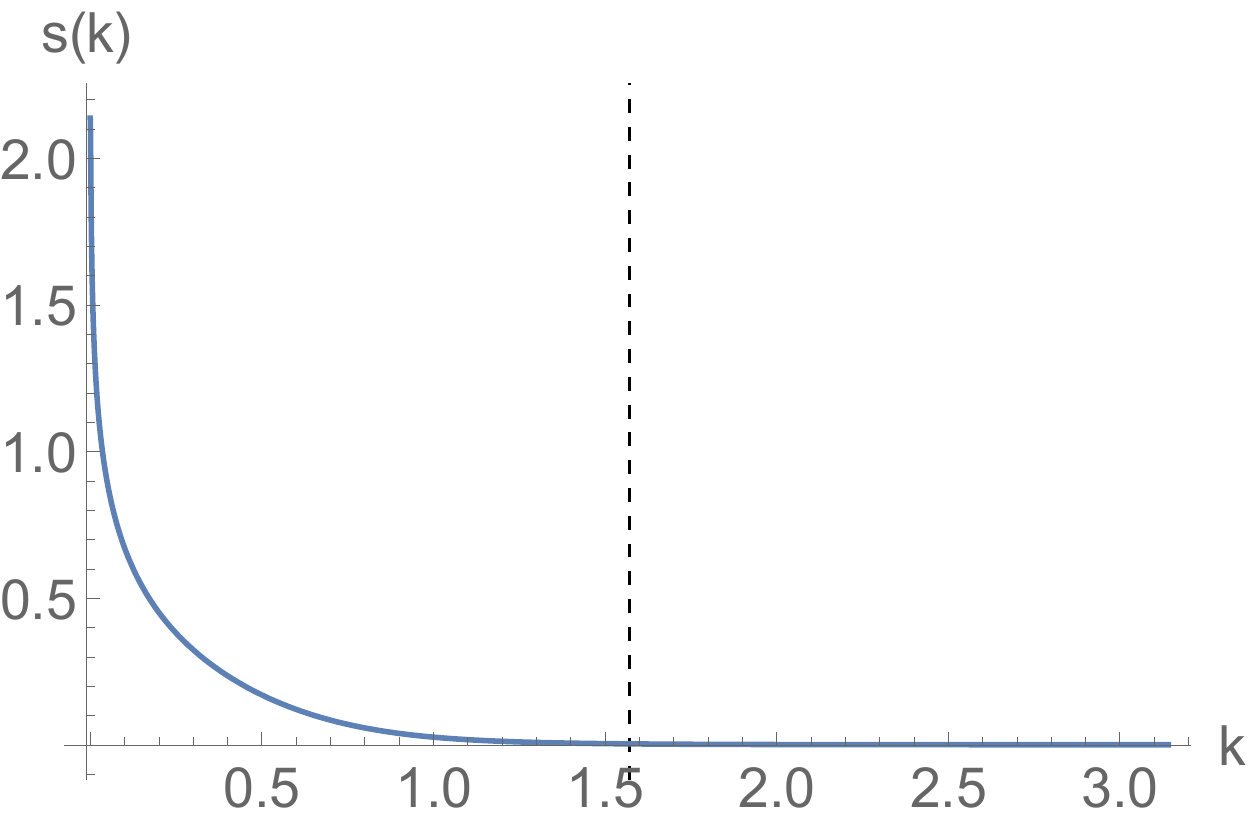} \label{fig14b}} 
      {\includegraphics[width=6 cm]{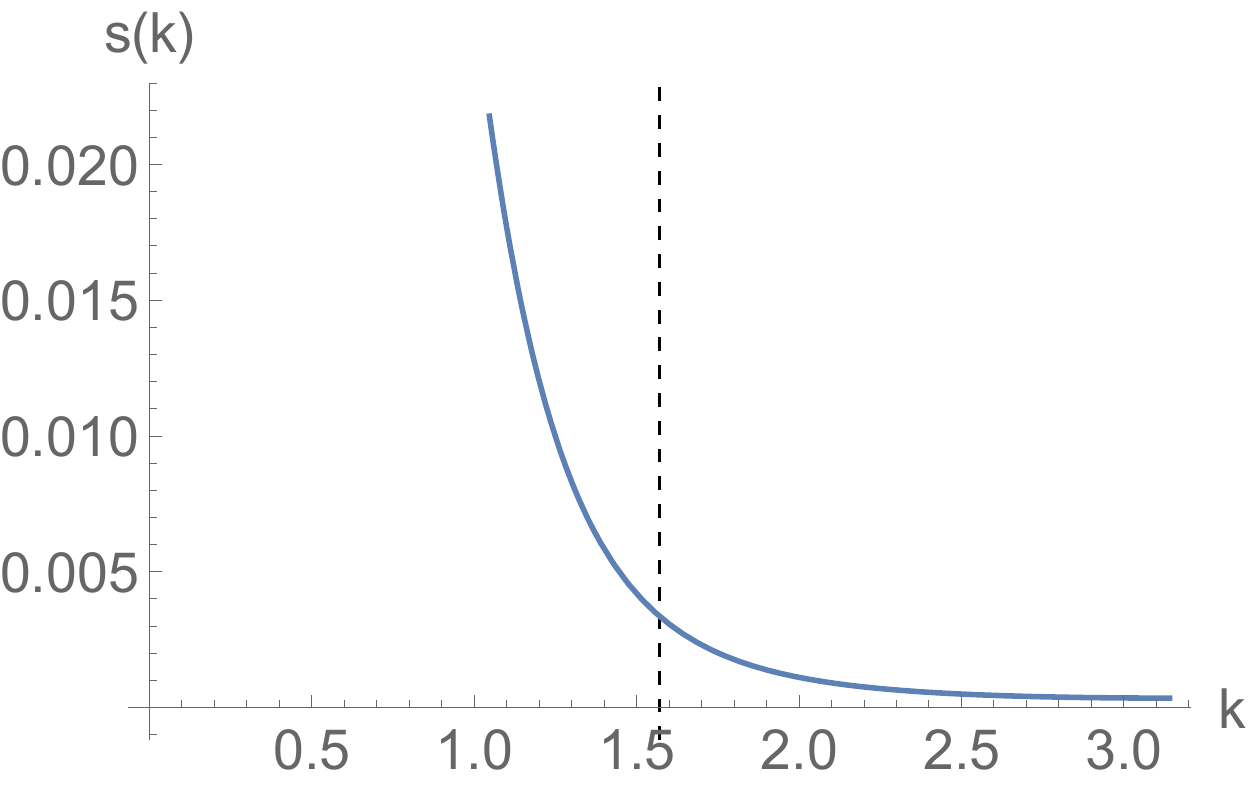} \label{}}}
           \caption{{The plot of} entropy density $s(k)$ for $z=2$ and $m_f=10^{-6}$.  The dashed lines represent the momentum $k^*$. }
  \label{skz11}
\end{figure}
\paragraph{$z > 2$ case} The qualitative feature of $v_k$ and $s(k)$ for $z>2$ are the same as the $z=2$ case. {The peak of  $v_k$ is more shifted to the right as $z$ increases.} Thus $t_c$ is delayed by the same reason. 
The general behavior of  $v_k$ can be understood by \eqref{gv}. If we take $m_f=0$ then, the dependence of {$\sin[k/2]$} will disappear only for $z=1$.

In short, in the quasiparticle picture the delayed $t_c$ can be explained by the small contribution of the fast quasiparticles to the entanglement entropy.

\section{Conclusions}\label{summary}
We have studied the time evolution of the entanglement entropy in {the free Lifshitz scalar theories} with the time-dependent mass by the correlator method on {1 dimensional} spacial lattice. 
The mass potentials are smooth functions {of} time (ECP or CCP), and the initial ground states evolve in time by the time-dependent Hamiltonians. 

Some important observations and comments from our computations are as follows. 
\begin{enumerate}
\item {At early times: For both ECP and CCP, the entanglement entropy is subsystem size independent. It can be understood intuitively by the quasiparticle picture {as shown in} Figure \ref{quasip}.}
\item The (intermediate) critical time: {From a naive application of the quasiparticle picture {\it only} by the fast quasiparticle, we expect the critical time $t_c$ for the significant $l$-dependence in the sudden quench to become $t_c \sim\frac{l}{2 |v_\textrm{max}|} $.} It works for $z=1$ {in the sudden quench}, however, {for the fast and slow ECP and the fast CCP with $z>1$,  we have found that} {it is possible 
\begin{equation}
 t_c \gg t_{\textrm{kz}} + \frac{l}{2 |v_\textrm{max}|} \,,
\end{equation}
where $t_{\textrm{kz}}$ is the Kibble-Zurek time, the time scale when the quasiparticle pairs {with $k \sim 0$} are created. For the Fast ECP and CCP case, $t_{\textrm{kz}} \sim 0$.}
{We have explained that,  by the quasiparticle formula in the sudden quench,}  it can be interpreted by the negligible contribution of the fast quasiparticles due to its small entropy density. Indeed, the dominant contribution comes from the slow quasiparticles. (For $z=1$ case, the entropy density is large for the fast quasiparticles, which is why the {\it only} fast particle approximation works for $z=1$.) {In addition, $t_c$ increases as the {subsystem sizes increase}. }
\item At late times: For the ECP, the entanglement entropy is slowly increasing. For the CCP with the final mass $m_f${,} the entanglement entropy is oscillating with a period $\sim \pi/m_f$, which can be understood by the dominant contribution by the slow quasiparticles with the momentum $k \sim 0$. 
\item $z$-dependence: The entanglement entropy increases as $z$ increases, which can be interpreted as the effect of the long-range interactions due to {the higher derivative $\partial^z_x$ in (\ref{H1}). }
\item {The time scale for the first local minimum of the entanglement entropy in the CCP case:  It is around $2\xi$ for the fast CCP and $2\xi_{\textrm{kz}}$ for the slow CCP {\it independently} of $z$. }
\end{enumerate}

Even though we {have found} that the quasiparticle picture is successful in understanding some of our results qualitatively for $z>1$ as well as $z=1$, note that it is a picture for the {\it sudden} quench. 
For a slow change of the mass potential or for a small final mass (see footnote \ref{f12}), there will be quantitative differences from the quasiparticle picture. Therefore, the property {items 1 and 2 can be explained well by the quasiparticle picture,} but it is not easy to determine precise $t_c$ by that picture. It is also not easy to determine the value of $t_c$ from our numerical correlator method; we first need to define some criteria for the significant deviation due to the {subsystem} size. Some of our results {cannot} be explained even qualitatively by that picture. For example, the nonlinear behavior of $\Delta S_A$ with large $z$ in Figure \ref{ecpf1} cannot be explained. It will be interesting to understand the $z$-dependence of this nonlinear behavior\footnote{{See also discussion of the $z$-dependence of entanglement entropy at early times in \cite{MohammadiMozaffar:2018vmk}.}} as well as more detailed understanding of item 4.  {Furthermore, it will be also interesting to consider the entanglement entropy at a limit $z\to\infty$ for understanding the effect of large $z$.}

Based on the argument in item 3, we may say that the entanglement entropy in our ECP does not oscillate at late times because $m_f \sim 0$. {If we considered the ECP with a finite $m_f$, we would have observed an oscillation.} It will be interesting to check this expectation by the correlator method. 
The last item 5 is very interesting since it shows a universal property independent of $z$. Even though our results are numerical, it seems very robust. We {do not} have a good understanding on it yet, and leave it as a future work.

\section*{Acknowledgement}
We would like to thank Kyoung-Bum Huh, Hyun-Sik Jeong, Chang-Woo Ji,  and  Ali Mollabashi for fruitful discussions.
The work of K.-Y. Kim, M.~Nishida, and M.-S.~Seo was supported by Basic Science Research Program through the National Research Foundation of Korea (NRF) funded by the Ministry of Science, ICT $\&$ Future Planning (NRF- 2017R1A2B4004810) and GIST Research Institute (GRI) grant funded by the GIST in 2019.
M.~Nozaki is supported by JSPS Grant-in-Aid for Scientific Research (Wakate) No.19K14724.
The work of M.~Nozaki is supported by RIKEN iTHEMS Program.
The work of M.~Nozaki and AT were the RIKEN Special Postdoctoral Researcher program.
 K.-Y. Kim, M.~Nishida, and M.-S.~Seo also would like to thank the APCTP (Asia-Pacific Center for Theoretical Physics) focus program,``Holography and geometry of quantum entanglement'' in Seoul, Korea for the hospitality during our visit, where part of this work was done.

\bibliography{HyunSikRefs}
\bibliographystyle{JHEP}

\providecommand{\href}[2]{#2}\begingroup\raggedright\endgroup

\end{document}